\documentclass[epj]{svjour}

\usepackage{amssymb}
\usepackage{amsmath}

\usepackage{graphicx}

\usepackage[numbers]{natbib}

\usepackage{hyperref}
\def\eq#1{\eqref{#1}}
\def\Eq#1{Eq.~\eqref{#1}}

\def\fig#1{Fig.~\ref{#1}}

\def\tab#1{Table~\ref{#1}}
\def\Sec#1{Sec.~\ref{#1}}

\graphicspath{
    {./figures/}
    {../figures/}
}

\newcommand{\tr}{{\textrm{tr}}}
\newcommand{\imag}{\textrm{i}}
\newcommand{\spatial}[1]{\vec{#1}}
\def\pslash{p\llap{/}}

\newcommand{\re}{{\textrm{Re}}}
\newcommand{\im}{{\textrm{Im}}}

\newcommand{\Tcep}{T_{CEP}}
\newcommand{\muBcep}{\mu_{B,CEP}}

\usepackage{xcolor}

\begin{document}

\title{Lee-Yang edge singularities in QCD via the Dyson-Schwinger Equations}


\author{
    Zi-Yan Wan\inst{1} \and Yi Lu\inst{1}
    \and Fei Gao\inst{2}\thanks{\emph{Corresponding author:} fei.gao@bit.edu.cn}
    \and Yu-xin Liu\inst{1}\inst{3}\thanks{yxliu@pku.edu.cn}
}

\institute{
    Department of Physics and State Key Laboratory of Nuclear Physics and Technology, Peking University, Beijing 100871, China
    \and School of Physics, Beijing Institute of Technology, 100081 Beijing, China
    \and Center for High Energy Physics, Peking University, Beijing 100871, China
}


\date{\today}

\abstract{
We take the Dyson-Schwinger Equation approach of QCD for the quark propagator at complex chemical potential to study the QCD phase transition. The phase transition line of the $(2+1)$-flavor QCD matter in the imaginary chemical potential region is computed via a simplified truncation scheme, whose curvature is found to be consistent with the one at real chemical potential. Moreover, the computation in the complex chemical potential plane allows us to determine the location of the Lee-Yang edge singularities. We show explicitly that the critical end point coincides with the Lee-Yang edge singularities on the real $\mu_{B}^{}$ axis. We also investigate the scaling behavior of the singularities and discuss the possibility of extrapolating the CEP from a certain range of chemical potential.
%
%
}

\maketitle

\begin{sloppypar}

\section{Introduction}

Many efforts have been made both experimentally and theoretically to explore the  phase structure of QCD  in the plane of temperature and chemical potential~\cite{Aarts:2023vsf,Hippert:2023bel,Huang:2023ogw,Fu:2022gou,Lovato:2022vgq,HADES:2019auv,Fischer:2018sdj,Luo:2017faz,Braun-Munzinger:2015hba,Schaefer:2008ax}.
At vanishing chemical potential, lattice QCD simulation has made great progresses and predicted a crossover at about $T=156.5\;$MeV~\cite{Borsanyi:2020fev,HotQCD:2018pds}.
However, due to the sign problem, it is difficult for lattice QCD to approach large chemical potential region.
In particular, the up to date computation  from functional QCD approaches~\cite{Gao:2020fbl,Gunkel:2021oya,Gao:2020qsj,Fu:2019hdw} and also effective theories~\cite{Hippert:2023bel,Cai:2022omk} have converging estimations that the  the critical end point (CEP) of chiral phase transition is located at a large chemical potential, $\mu_{B}^{}\approx 600\,\textrm{MeV}$.
Therefore, people  appeal to the imaginary chemical potential where there is no sign problem and   can offer more information about the phase structure and CEP.

It has been known that the imaginary chemical potential   offers a way of extracting the curvature of the phase transition line of QCD after assuming the line is continuous from imaginary to real chemical potential.
%
%
And the analytical structure of the thermodynamic quantities  determines the phase structure.
In detail,  the edge singularities, i.e. the Lee-Yang zeroes,  are  located in the complex plane of the  chemical potential and at second or first order phase transition, the singularities  reach the real axis~\cite{Yang:1952be,Lee:1952ig}.
Consequently, the Lee-Yang zeroes also determine the convergence radius of the expansion for finite chemical potential applied in lattice QCD simulations~\cite{Clarke:2023noy,Schmidt:2022ogw,Singh:2021pog}. Moreover, there are more interesting features in the imaginary chemical potential region. For instance, there exists the Roberge-Weiss phase transition, which can be connected to the confinement due to its relation with $Z(N_{c})$ symmetry \cite{Roberge:1986mm,Fischer:2014vxa}. The imaginary chemical potential dependence of chiral condensate, i.e. the dual condensate \cite{Fischer:2009wc,Fischer:2011mz}, is thus also related to the confinement.

Plenty of novel phenomena rise up in the complex chemical potential region of QCD.
However, the related physics have not yet been fully investigated theoretically under the QCD approaches.
In particular, the Dyson-Schwinger equations (DSEs) approach has been shown successful of incorporating both the dynamical chiral symmetry breaking (DCSB) and the confinement physics of non-perturbative QCD~\cite{Fischer:2018sdj,Alkofer:2000wg,Roberts:2007ji}, which is also well developed for the application on exploring the QCD phase structure in recent years~\cite{Gao:2020fbl,Gunkel:2021oya,Gao:2020qsj}.
We then present for the first time a DSEs study on the  Lee-Yang edge (LYE) singularities and its trajectory on the complex $\mu_{B}^{}$ plane.
Most importantly, we show directly the close relation between the LYE singularities and the CEP location.

In specific, we provide both an estimation of the CEP location, as well as a direct computation of the trajectory of the LYE singularities.
We provide an estimate of the critical region by studying the extrapolation of the trajectory towards the actual CEP location.
Our studies of the scaling behavior of LYE singularities give a clear picture on   the QCD phase structure, with connections to various different aspects such as the critical scaling, the Columbia plot, the convergence radius of the equation of state, etc. In particular,  We illustrate that  the trajectory of LYE singularities satisfies the CEP scaling in a large range of temperature. We then show that the LYE singularities offers a new way of estimating the location of CEP, and our estimation is close to the onset regime of CEP at $T \simeq 110\,\textrm{MeV}$ and $\mu_B \simeq 600\,\textrm{MeV}$ obtained from extracted information from LYE singularities in combination with  the  current best truncation schemes of functional QCD approaches.

The remainder of this paper is organised as follows.
In \Sec{model}, we present the  framework of Dyson-Schwinger equations we implemented here.
In \Sec{LYE}, we deliver the results of Lee-Yang edge singularities, and in \Sec{anlys} we make some detailed analysis for the QCD phase structure based on the Lee-Yang edge singularities.
In \Sec{sec:modelparam}, we analyse the dependence of the result on the parameters in the truncation scheme.  In \Sec{sum}, we briefly summarize the main results and make some discussions.

\section{The DSEs approach and the truncation scheme}\label{model}

\begin{figure}[b]
\centering
\includegraphics[width=0.45\textwidth]{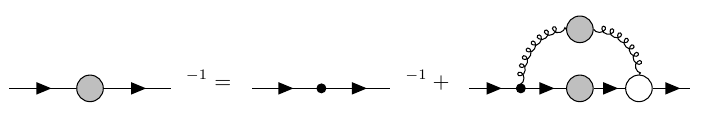}%
\caption{A schematic plot for the quark DSE. The line with black dot stands for the bare quark propagator, and the line with solid blobs stand for the full propagators and the circle for the interaction vertice.}
    \label{quarkDSE}
\end{figure}

In this paper, we take the DSEs approach to study the QCD phase structure at finite temperature $T$ and chemical potential $\mu_{B}^{}$.
Our main focus is to solve the quark DSE, schematically shown in \fig{quarkDSE}.
At finite temperature $T$ and quark chemical potential $\mu$, the $O(4)$ symmetry of the momentum $p$ is broken to $O(3)$ $(\spatial{\,p},\omega_n)$, where $\omega_{n} = 2\pi T(2n+1)$ is the Matsubara frequency for fermion. The dressed quark propagator is written as:
\begin{equation}\label{eqS}
 \left[S(p)\right]^{-1}=\imag \tilde{\omega}_{n} \gamma_{4} C(\spatial{\,p},\tilde{\omega}_{n}) + \imag \spatial{\,p} \spatial{\gamma} A(\spatial{\,p},\tilde{\omega}_{n}) + B(\spatial{\,p},\tilde{\omega}_{n}) \, ,
 \end{equation}
with abbreviation $\tilde{\omega}_n=\omega_n+\imag \mu$, $A,B,C$ are scalar functions that need to be determined.
Together with the dressed gluon propagator $D_{\mu \nu }$ and the gluon-quark interaction vertex $\Gamma_{\mu}$, \fig{quarkDSE} can be expressed as:
\begin{equation}\label{eq1}
\begin{split}
& \left[S(p)\right]^{-1} = \left[S_0(p)\right]^{-1} + \Sigma(p), \\
& \Sigma(p) = C_{F} Z_{1 } g^{2} T \sum_{n} \int \frac{d^{3} \spatial{\,q}}{(2 \pi)^{3}} \gamma_{\mu} S(q) \Gamma_{\nu}(p, q ; k) D_{\mu \nu}(k) \, ,
\end{split}
\end{equation}
with the renormalized bare quark propagator
\begin{equation}\label{eq:Sq0}
 \left[S_{0}(p)\right]^{-1} = Z_{2} \imag \pslash + Z_{2}Z_{m} m_{f} \, ,
\end{equation}
where $Z_{2},Z_{1},Z_{m}$ are quark wave function, vertex, and quark mass renormalization constants, and $C_{F} = \frac{N_{C}^{2} -1}{2N_{C}}$ is  the expectation value of  the Casimir operator $\hat{C}_{2,SU(N_{C})}$; $g^{2} = 4\pi \alpha(\zeta)$ is the coupling constant which is dependent on the renormalization point $\zeta$.
The renormalisation condition of the quark DSE is:
\begin{equation}\label{eq:renorm}
\begin{aligned}
& Z_{2} = 1 - \Sigma_{d}(\zeta^{2}), \qquad Z_{2}Z_{m} = 1 - \frac{\Sigma_{s}(\zeta^{2})}{m_{f}} , \\
& \textrm{with} \quad \Sigma(p^{2}) \equiv \imag \pslash \Sigma_{d}(p^{2}) + \Sigma_{s}(p^{2}) .
\end{aligned}
\end{equation}
We solve the quark DSE with a renormalization point at $\zeta=40\,\textrm{GeV}$, with the corresponding running coupling taken as $\alpha(\zeta)=0.1286$,
and the average current mass of $u$, $d$ quark $m_{l} = (m_{u} + m_{d})/2 = 2.09\,\textrm{MeV}$.

\begin{figure}[t]
\centering
\includegraphics[width=0.85\linewidth]{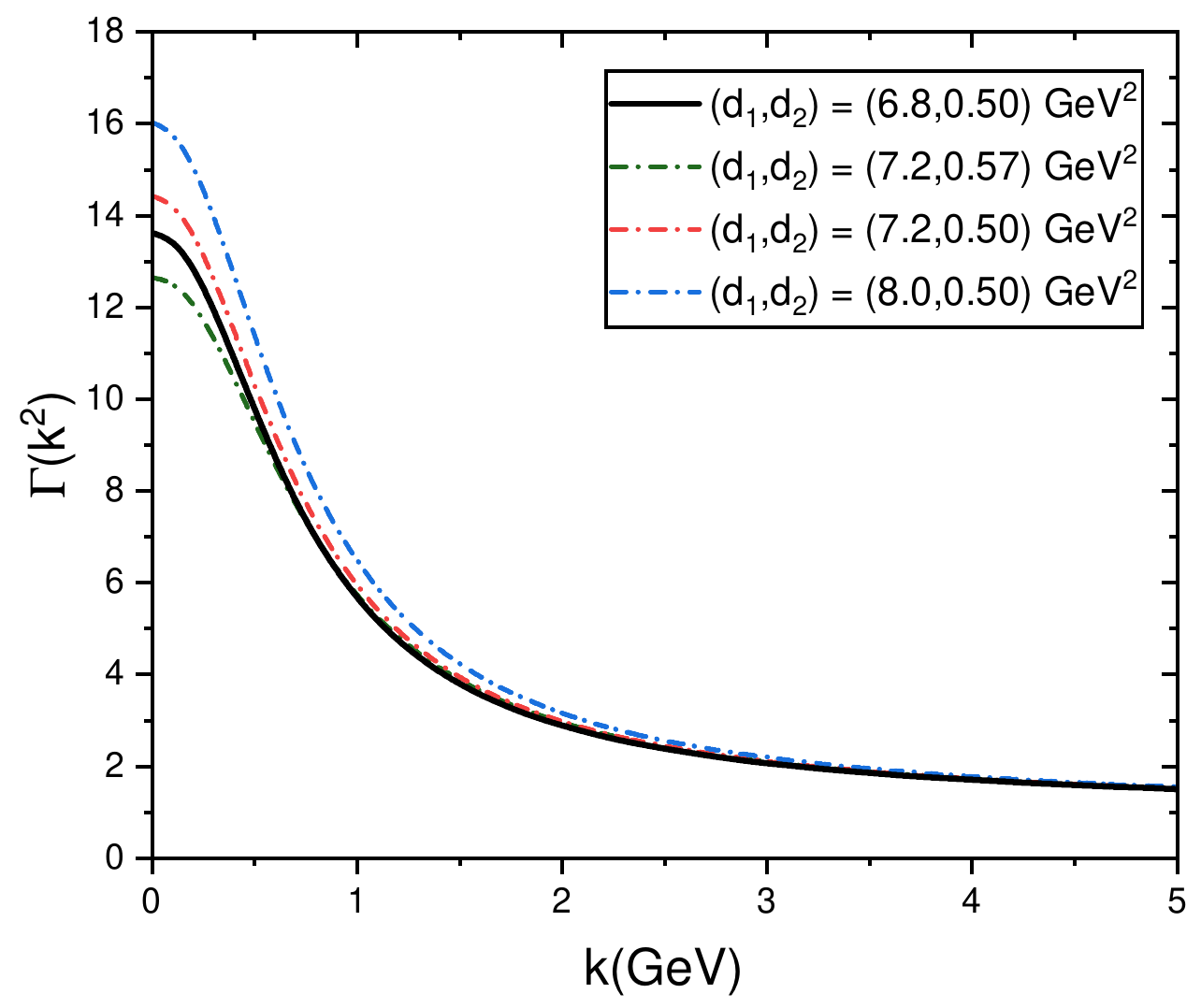}
\caption{The running behaviour of the quark-gluon vertex function $\Gamma(k^2)$ in the truncation scheme \Eq{vertex}. A comparison of the function under different choices of the parameter $(d_1,d_2)$ is also shown, including the choice taken from Ref.~\cite{Fischer:2014ata} as $(6.8,0.5)\,\mathrm{GeV}^2$ (solid black curve) and for other values (colored dash-dotted curves, to be investigated in \Sec{sec:modelparam}).
}\label{fig:vtx}
\end{figure}

For the truncation of the quark-gluon interaction vertex, we follow the Ans\"{a}tze used in Refs. \cite{Gunkel:2021oya,Fischer:2014ata}, which follows from the Slavnov Taylor identity (STI) with a parametrised form for the non-Abelian part of the vertex $\Gamma$ as:
%
%
 \begin{equation}\label{vertex}
 	\begin{aligned}
 		\Gamma_{\mu}(p,q;k) = & \gamma_{\mu}\cdot\Gamma(k^{2})\cdot \\
            & \left(\delta_{\mu,4}\frac{C(p)+C(q)}{2}+\delta_{\mu,i}\frac{A(p)+A(q)}{2}\right), \\
 	\Gamma\left(k^{2}\right) = & \frac{d_{1}}{d_{2}\!+\! k^{2}} \! + \! \frac{k^{2}}{\Lambda^{2}\! + \! k^{2}}\! \Bigg{(}\! \frac{\beta_{0}\alpha(\zeta)\ln\left[k^{2}/\Lambda^{2}+1\right]}{4\pi}\! \Bigg{)}^{2\delta}.
 	\end{aligned}
 \end{equation}
%
%
where the anomalous dimension $\beta_{0} =(11N_{c} - 2N_{f})/3$, $\delta=-9N_{c}/(44N_{c} - 8N_{f})$.
For the parameters we will first follow Ref.~\cite{Fischer:2014ata} (set $B_{2+1}$) to take $\alpha(\zeta)=0.3$, $\Lambda = 1.4\,\textrm{GeV}$, $d_{1}=6.8\,\textrm{GeV}^{2}$ and $d_{2} = 0.5\,\textrm{GeV}^{2}$, which are determined by solving the Bethe-Salpeter equation for the 2+1-flavor case.
The running behavior of such a $\Gamma(k^2)$ is shown as the solid black curve in \fig{fig:vtx}. 
Moreover, we also show the dependence of $d_1$ and $d_2$ on $\Gamma(k^2)$ by varying the value of the two parameters, which is also shown in \fig{fig:vtx}. 
A more detailed analysis on the parameter dependence for $d_1$ and $d_2$ will be given in \Sec{sec:modelparam}.

The (2+1)-flavor gluon propagator takes the input from the called ``functional-lattice" gluon propagator in Ref.~\cite{Gao:2021wun}. In the vacuum, the Landau gauge gluon propagator $D_{\mu \nu }$ is expressed as:
 \begin{equation}
 D_{\mu \nu}(p)=D(p) P_{\mu \nu }, \quad \text{  with }P_{\mu \nu} = \delta_{\mu \nu} - \frac{p_{\mu} p_{\nu}}{p^{2}},
 \end{equation}
 and $D(p)$ the gluon dressing function, which includes both the lattice results at small momentum and the FRG results at large momentum:
  \begin{equation}
  		\begin{aligned}
  			&D(p)=\frac{\left(a^{2}+p^{2}\right)/\left(b^{2}+p^{2}\right)}{M^{2}\left(p^{2}\right)+p^{2}\left[1+c\ln\left(d^{2}p^{2}+e^{2}M^{2}\left(p^{2}\right)\right)\right]^{\gamma}},\\
  	&\text{ with }	M^{2}(p)=\frac{f^{4}}{g^{2}+p^{2}},
  	\end{aligned}
 \end{equation}
  where the anomalous dimension $\gamma = \frac{13 - 4/3N_{f} }{\, 22 - 4/3N_{f} \, }$, along with the other parameters given by $\{a,b,c,d,e\}=\left\{ 1\,\mathrm{GeV},\, 0.735\,\mathrm{GeV},\, 0.12,\, 0.0257\,\mathrm{GeV}^{-1},\, 0.081\,\mathrm{GeV}^{-1}\right\}$, together with $\{f,g\}=\{0.65\,\mathrm{GeV},\, 0.87\,\mathrm{GeV}\}$.

At finite temperature $T$ and baryon chemical potential $\mu_{B}^{}$, the gluon propagator is split into the longitudinal ($L$) and transversal ($T$) part with respect to the heat bath by the corresponding projectors $P_{\mu \nu }^{T,L}$.
  \begin{equation}
  	\begin{aligned}
  		P^{T}_{\mu \nu }=&(1-\delta_{\mu 4})(1-\delta_{\nu 4})P_{\mu \nu},\\
  		P^{L}_{\mu \nu }=& P_{\mu \nu} - P^{T}_{\mu \nu},
  	\end{aligned}
  \end{equation}

Also, the finite $T$ and $\mu_{B}^{}$ effect on the gluon propagator is incorporated into the longitudinal part $D^{L}$ by the thermal mass,
which can be taken as that provided by the one-loop hard thermal loop (HTL) calculations~\cite{Vija:1994is,Haque:2012my}:
\begin{equation}
D_{\mu \nu}=D^{T}(\Omega ^{2}_{nm},\spatial{k}) P^{T}_{\mu \nu} + D^{L}(\Omega ^{2}_{nm} + m_{g}^{2},\spatial{k}\,)P^{L}_{\mu \nu}.
\end{equation}
\begin{equation}\label{eq:HTLmass}
    m_{g}^{2} = \frac{g^{2}}{3}\left[(N_{c}+\frac{N_{f}}{2})T^{2} + \frac{3N_{f}}{2\pi^{2}}\mu^{2} \right].
\end{equation}
with $\mu = \mu_{B}^{}/3$, i.e. the 3-flavor degenerate case.
As for the case of a complex chemical potential with $\im\,\mu=\phi \pi T$ satisfying $\phi\in[-1,1]$, the thermal mass $m_{g}^{2}$ is extended to complex plane according to the analytical continuation of \Eq{eq:HTLmass}.
In practice, due to the small effect of $\im\,\mu$ on the gluon sector, we maintain only the real part of $m_{g}^{2}$ for convenience.

\begin{figure}
\centering
    \includegraphics[width=0.85\linewidth]{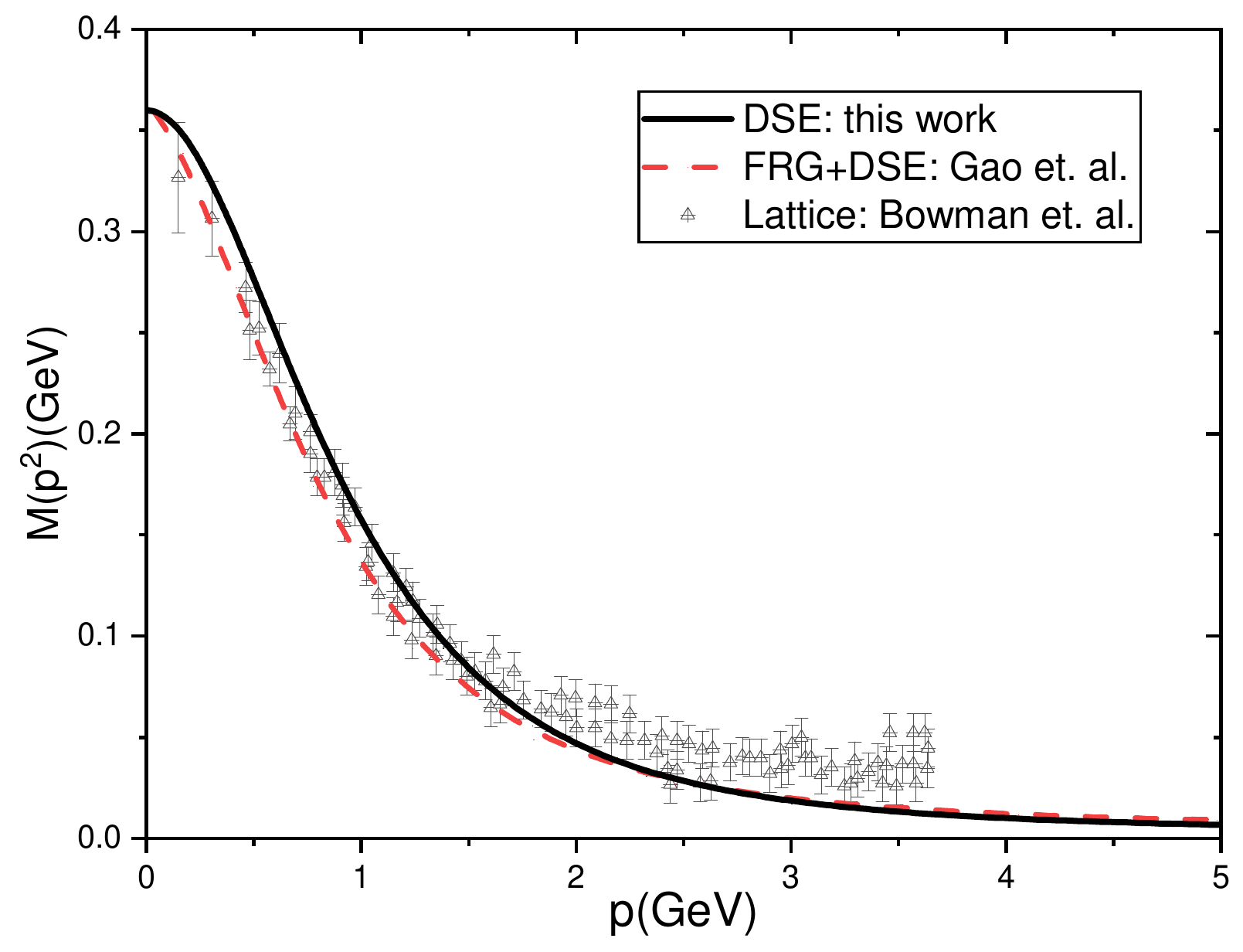}
    \caption{Calculated quark mass function in the vacuum, and the comparison with the lattice QCD result \cite{Bowman:2005vx} and the fully coupled functional QCD result~\cite{Gao:2021wun}.
    }
    \label{M-dynamical}
\end{figure}

First, we solve the quark DSE in \Eq{eq1} self-consistently in the vacuum, and obtain the quark mass function $M(p)=B(p)/A(p)$ as shown in \fig{M-dynamical}.
which is consistent with the results from the lattice QCD simulation~\cite{Bowman:2005vx} and the fRG-DSE calculation~\cite{Gao:2021wun}.
Moreover, we focus on the quark condensate $\langle \bar{\psi} \psi \rangle$ which can be calculated from the quark propagator:
\begin{equation}\label{con}
	\langle \bar{\psi} \psi \rangle = Z_{2}Z_{m} N_{c} T \sum_{n} \int \frac{d^{3} \spatial{\,p}}{(2\pi)^{3}}\tr[S(\tilde{\omega}_{n},\spatial{\,p})] \, ,
\end{equation}
For a non-zero bare quark mass $m_{l}$, \Eq{con} will be linearly divergent.
Here, one of the regularization scheme is that eliminates the divergent part via the derivative form as done in Refs.~\cite{Gao:2021wun,Gao:2016qkh}:
\begin{equation}\label{chi1}
	\langle \bar{\psi} \psi \rangle_{\mathrm{reg}} = \langle \bar{\psi} \psi \rangle - m\frac{\partial}{\partial m}\langle \bar{\psi} \psi \rangle .
\end{equation}
Another choice is the reduced condensate, defined from the strange quark condensate in Refs.~\cite{Fischer:2014ata,Lu:2023mkn}
\begin{equation}\label{chi2}
	\langle \bar{\psi} \psi \rangle_{\mathrm{red}}=\langle \bar{\psi} \psi \rangle^{l} - \frac{m_{l}}{m_{s}} \langle \bar{\psi} \psi \rangle^{s} \, ,
\end{equation}
with $m_{s} = 27\, m_{l} = 56.43\,\textrm{MeV}$. 
With the quark propagator determined by solving \Eq{eq1},
we obtained the regularized condensate from \Eq{chi1} as $\langle \bar{\psi} \psi \rangle_{\mathrm{reg}} = (333.6\,\textrm{MeV})^3$, and the reduced condensate from \Eq{chi2} as $(335.8\,\textrm{MeV})^{3}$.
The obtained regularized quark condensate satisfies excellently the Gell-Mann-Oakes-Renner Relation:
\begin{equation}
	f_{\pi} ^{2} m^{2}_{\pi} =2 m^{\zeta} \langle \bar{\psi} \psi \rangle_{\zeta}\, ,
\end{equation}
with the vacuum pion decay constant $f_{\pi} =89\,\textrm{MeV}$ and the pion mass $m_\pi=139\,\textrm{MeV}$. \\

\begin{figure}
    \centering
	\includegraphics[width=0.85\linewidth]{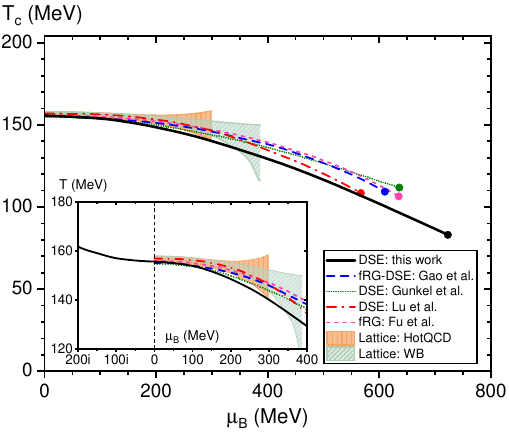}
	\caption{The obtained phase diagram of the chiral phase transition identified with the susceptibility \Eq{eq:chim}, compared with the lattice QCD \cite{Borsanyi:2020fev,HotQCD:2018pds} and other functional QCD \cite{Gao:2020fbl,Gunkel:2021oya,Fu:2019hdw,Lu:2023mkn} results. The solid dots denote the positions of the CEP. A comparison between pure imaginary and real $\mu_{B}^{}$ is also shown in the sub-figure. }
	\label{ptd}
\end{figure}

At finite temperature and chemical potential, one can make use of the maximum of the light-quark chiral susceptibility $\chi_{m}^{l}$ to determine the pseudo-critical temperatures, which is defined from the chiral condensate as
\begin{equation}\label{eq:chim}
	\chi_{m}^{l} = \frac{\partial \langle \bar{\psi} \psi \rangle_{\mathrm{reg}}}{\partial m_{l}} \, ,
\end{equation}
For the reduced condensate in \Eq{chi2}, the expression of $\chi_{m}$ can be reduced to ${\partial \langle \bar{\psi} \psi \rangle^l}/{\partial m_l}$ due to the small variations on the strange condensate according to $m_{l}$.
Based on this, we take a further approximation of \Eq{eq:chim} as:
\begin{equation}\label{chi}
	\chi=\frac{\partial B_{0}}{\partial m_{l}},
\end{equation}
where $B_{0}$ is the value of $B$ at the first Matsubara frequency and zero three-momentum $B(\tilde{\omega}_{0},\boldsymbol{0})$.
Such an approximation has been confirmed to be efficient in, e.g., Ref. \cite{Gao:2016qkh}, with only a few MeV difference of the extracted pseudo-critical temperatures.
Then we obtain the phase transition line as displayed in \fig{ptd}.

At zero chemical potential,  our obtained pseudo-critical temperature is $T_{c}=155.7\,\textrm{MeV}$, which agrees excellently with the up-to-date lattice QCD results~\cite{Borsanyi:2020fev,HotQCD:2018pds} and the functional QCD results~\cite{Gao:2020fbl,Gunkel:2021oya,Fu:2019hdw,Fischer:2014ata,Lu:2023mkn}.
At larger chemical potential and lower temperature, the chiral susceptibility as a function of $T$ and $\mu$  becomes sharper and sharper, and the critical temperature decreases.
Such a quantitative nature is embedded in the curvature of the phase transition line:
\begin{equation}\label{eq:PTline}
  \frac{T_{c}(\mu_{B})}{T_{c}(0)} =  1 - \kappa_{2}^{B} \left( \frac{\mu_{B}}{T_{c}(0)} \right)^{2} + \kappa_{4}^{B} \left( \frac{\mu_{B}}{T_{c}(0)} \right)^{4} + \cdots \, .
\end{equation}
The curvature $\kappa_{2}^{B}$ is extracted from a region of
$\mu_{B}^{} \in (0,360)\,\textrm{MeV}$ with a result $\kappa_{2}^{B} = 0.026$.
We also check for the pure imaginary chemical potential in a range of $\mu_{B}^{2} \in (-210^{2},0)\,\textrm{MeV}^{2}$, and obtain the corresponding curvature $\kappa_{2}^{B} = 0.024$. 
The phase transition line at real and pure imaginary $\mu_B$ is also displayed together in \fig{ptd}, which shows that the continuity is valid at zero $\mu_{B}^{}$, in agreement with what is found in Ref.~\cite{Bernhardt:2023ezo}.
Eventually, the susceptibility is divergent at the location of CEP, which is found at $(T_{CEP}, \mu_{B,CEP}^{}) = (83\,\textrm{MeV},723\,\textrm{MeV})$.
Compared with a similar truncation scheme as in Ref.~\cite{Fischer:2014ata}, one can notice that the $T_{CEP}^{}$ is almost the same, but the $\mu_{B,CEP}^{}$ is larger due to a different treatment on the gluon sector.



 \section{The Lee Yang edge singularities}\label{LYE}

In Refs.~\cite{Yang:1952be,Lee:1952ig}, Lee and Yang have revealed the relation between the phase transition and zeroes of the partition functions $Z$ on the complex plane of the $\mu$: the Lee-Yang Zeroes.
These zeroes form branch cuts on the $\mu$ plane, and the edge branch points are known as the Lee-Yang edge (LYE) singularities.
In short, when the LYE singularities reach the $\mu$ real axis, the singularities correspond to the critical end point.
While the branch cut crosses the real axis, the first order phase transition occurs.
These can be well understood due to the properties of pressure $P= \frac{1}{V} T \ln Z$ around the Lee-Yang Zeroes.
Briefly speaking, the first derivative of $P$ is discontinuous crossing the branch cut, and continuous but non-analytical at the LYE singularities.
This suggests that the same criterion, i.e. the susceptibility $\chi$ in \Eq{eq:chim} for the phase transition line introduced in \Sec{model} can also be applied to determine the location of LYE singularities, when the chemical potential is extended to the complex values.

\begin{figure}[t]
\centering
\includegraphics[width=0.5\textwidth]{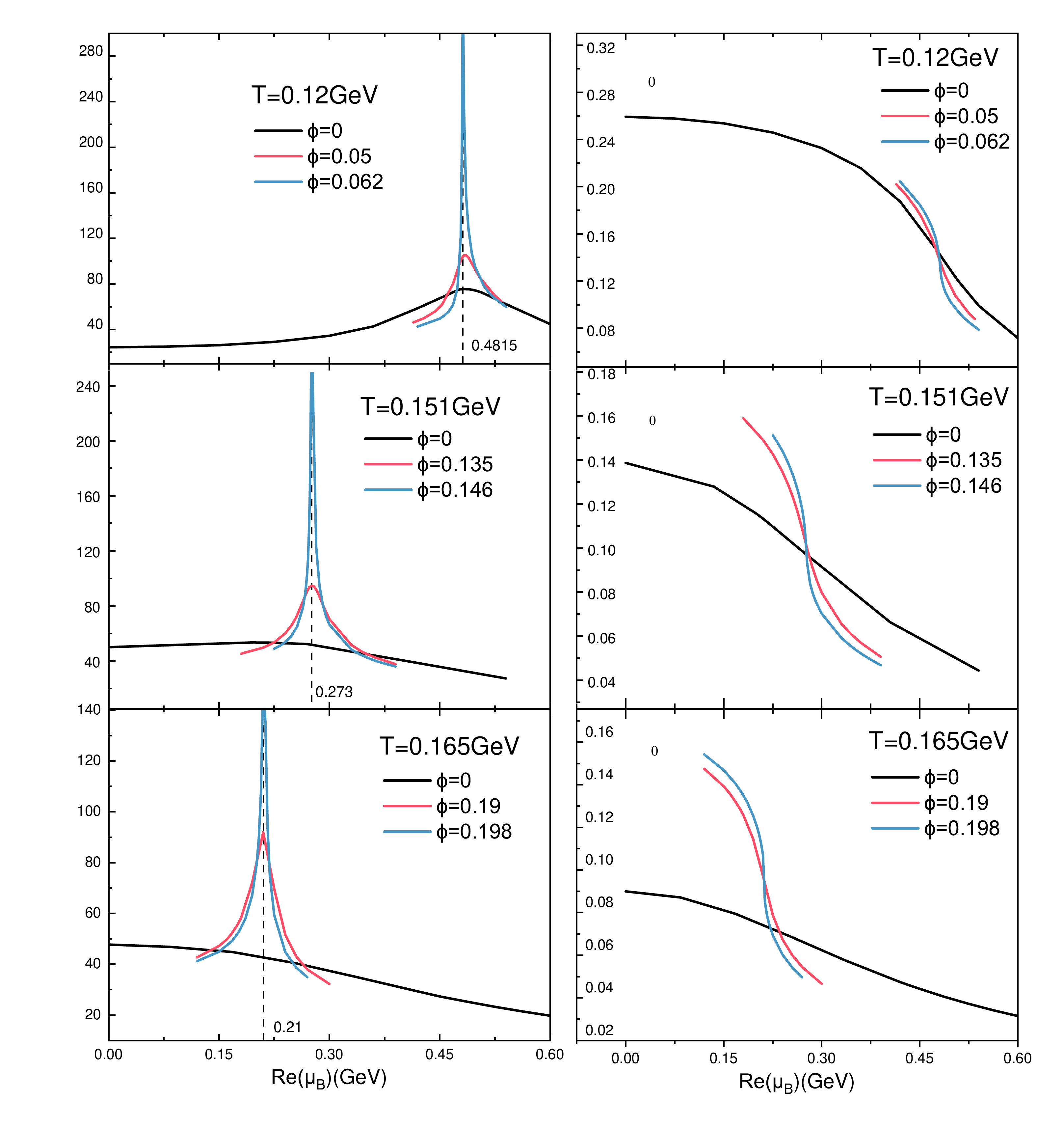}
\caption{The obtained complex chemical potential dependence of chiral susceptibility $\chi$ in \Eq{chi}  (left panels) and the mass  $M_{0}$ (right panels) at temperatures $T=120\,\textrm{MeV}$ (the first row), $T=151\,\textrm{MeV}$ (the second row), $T=165\,\textrm{MeV}$ (the third row). The singularities which are marked by the dashed vertical lines correspond to the critical chemical potentials.}
\label{LYSF}
\end{figure}

\begin{figure}[t]
\centering
\includegraphics[width=0.42\textwidth]{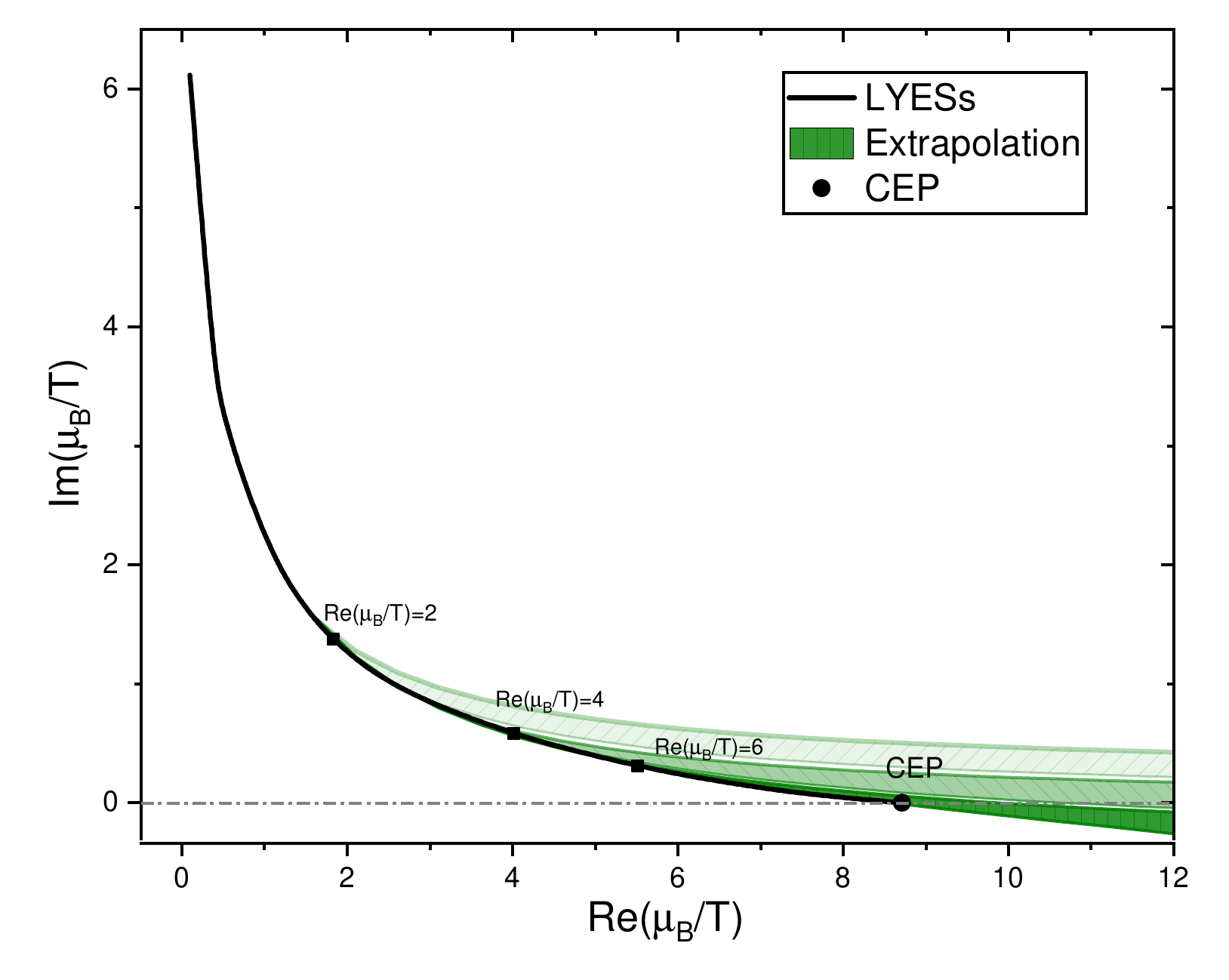}
\vspace*{-2mm}
\caption{The obtained trajectory of the LYE singularities on the $\re (\mu_B/T)$ -- $\im (\mu_B/T)$ plane, together with the CEP which is marked on the $\im (\mu_B/T)=0$ axis. The extrapolation analysis of the trajectory using data within $\re(\mu_B/T) \in (0,2)$, $(0,4)$ and $(0,6)$ are also shown with the error estimates, which are displayed as the light, medium and dark colored bands, respectively. }
\label{LYES_T}
\end{figure}

In specific, we scan the $\chi$ over the $T$--$\re\,\mu$ plane at a given imaginary chemical potential $\im\,\mu=\phi \pi T$ for $\phi\in(0,1)$, and the obtained result is illustrated in \fig{LYSF}.
It can be seen that when $T > T_{CEP}$, there exist apparent singularities on the complex $\mu$ plane.
For example, at $T=120\,\textrm{MeV}$, the susceptibility $\chi$ is continuous along $\re(\mu)$ at $\phi = 0$.
As $\phi$ approaches the critical value $\phi _{c} =0.062$, $M_{0}$ becomes steeper with a sharper peak of $\chi$ and finally reaches the singularity at $\re\,{\mu_{LYE}^{}}=481.5\,\textrm{MeV}$.
For a higher temperature, e.g. $T=150\,\textrm{MeV}$, one can observe a similar behavior of $\chi$.
Also, at $T > T_{c}$ the singularity still exists in the complex $\mu_{B}^{}$ plane.
The edge of such a singularity $\mu_{LYE}^{}$ is scanned in a wide range of temperature to yield a trajectory of the LYE singularities, the obtained results are displayed in \fig{LYES_T}.
Our result shows apparently that the trajectory crosses the real axis of $\mu_{B}^{}$ exactly at $T = T_{CEP}$, which implies a strong connection between the QCD critical end point and the LYE singularities.
%

We then provide an estimate on the critical region of the CEP via an extrapolation study of the obtained $\mu_{LYE}^{}$ data.
In specific, we consider the data within several small chemical potential regions and carry out the extrapolation with the convergent Pad\'{e} polynomials (in $N=n\,m,\, n\in[1,2],\ m\in[n,3]$) towards a higher $\re\,\mu_{B}^{}$ to compare with the calculated trajectory and also the CEP position.
The regions we take are $\re(\mu_{B}^{}/T) \in [0,2]$, $[0,4]$ and $[0,6]$ and the corresponding analyses are shown in \fig{LYES_T}.
It is found that the quality of the extrapolation becomes better with an extending range of $\re(\mu_{B}/T)$ being included. Within $\re(\mu_{B}/T) \in [0,2]$ where the lattice extrapolation approach is still controllable, the extrapolated position of the CEP has a systematic error of more than $30\%$. 
It is also estimated that a knowledge of $\re(\mu_{B}/T)$ with the range corresponds to $\Delta T = |T/\Tcep - 1|\sim 20\%$ and $\Delta \mu = |\mu/\muBcep - 1|\sim 15\%$, is required in order to get an error control of 10\% for the extrapolated CEP.
In short, additional information at large $\re\,\mu_{B}^{}$ is still required in order to provide a precise determination of the CEP position.

\section{Analysing the trajectory of the LYE singularities}\label{anlys}
	
In the vicinity of the LYE singularities where the second-order phase transition occurs, the correlation length tends to be divergent,
and the critical behavior is only determined by the symmetry and dimension of the system which can be classified into various universality classes.
Then one can describe the critical behavior including the location of LYE singularities of a specific system by universal quantities and the non-universal relevant thermodynamic parameters.
For the Ising model and $O(N)$ model,  expressed by the scaling field $z$, the singularities locate at $z=t/h^{1/\beta \delta}=|z_c|e^{i\pi /2\beta\delta}$ \cite{Basar:2021hdf,Zinn-Justin:1989rgp},
where $t$ and $h$ are the general scaling variables with $t$ being  the reduced temperature,  and $h$ being the symmetry breaking field. 
The $|z_{c}|$, $\beta$ and $\delta$ are universal scaling parameters which can be determined in the universality class analysis, and some of which are listed in \tab{uni_para}.

\begin{table}[b]
\centering
\caption{The critical exponents $\beta $,  $\delta$ and the location of Lee Yang Edge Singularities $|z_{c}|$ for different universality classes ~\cite{Connelly:2020gwa,Johnson:2022cqv,Rennecke:2022ohx,Kos:2016ysd}.}
     \begin{tabular}{c|c|c|c}
\hline \hline
           & $\beta$ & $\delta$ & $|z_{c}|$
           \\
          \hline
$ 3d \,\, Z(2)$&0.33&4.79&2.43\\
          \hline
          $ 3d  \,\,  O(4)$ & ~~~~$0.38$~~~~ & ~~~~$4.82$~~~~ & ~~~~$1.69$~~~~ \\
         \hline
          Mean field & 0.5 & 3 & 1.89\\  \hline \hline
         \end{tabular}
 \label{uni_para}
\end{table}

In the chiral limit, the QCD Lagrangian has $SU_R(2) \times  SU_L(2)\cong O(4)$ symmetry. Therefore, for a small quark mass, the critical behavior is expected to be described by the 3D $O(4)$ universality class. According to Refs.~\cite{Kaczmarek:2011zz,Mukherjee:2019eou},
%
%
the thermodynamic parameters $T$, $\mu_{B}^{}$ and with the light-to-strange quark mass ratio $m_{l}/m_{s}$ can be mapped onto the scaling variables   $t$ and $h$  at zero chemical potential
 as:
\begin{equation}\label{th1}
	\begin{aligned}
	t=&\, t_{0}^{-1}\left[\frac{T - T_{c}^{0}}{T_{c}^{0}} + \kappa_{2}^{B} \left(\frac{\mu_{B}^{}}{T_{c}^{0}} \right)^{2} \right] \, ,\\
	h=& \, h_{0}^{-1}\frac{m_{l}}{m_{s}} \, ,
	\end{aligned}
\end{equation}
from which one can derive the expression for the location of the singularities at specific temperature~\cite{Mukherjee:2019eou}:
\begin{equation}\label{O4sca}
	  \left(\mu_{LYE}^{} \right)^{2} = \frac{(T_{c}^{0})^{2}}{\kappa_{2}^{B}}\left[\frac{|z_{c}|}{z_{0}}e^{i\pi /2\beta\delta} \left(\frac{m_{l}}{m_{s}}\right)^{\frac{1}{\beta \delta}} - \frac{T - T^{0}_{c}}{T^{0}_{c}} \right] \, ,
\end{equation}
where $\kappa_{2}^{B}$ is the curvature of the phase transition line and $T_{c}^{0}$ the phase transition temperature at zero chemical potential in chiral limit.
In turn, the singularity corresponds to the convergence radius of the expansion with respect to the chemical potential.
As determined in the last section, we have $\kappa_{2}^{B} = 0.026$ and $m_{l}/m_{s} = 1/27$.
The remaining parameters are $T_{c}^{0}$ and $z_{0}$, the former of which can be interpreted as the critical temperature in the chiral limit.
With our truncation scheme in \Eq{vertex}, such critical temperature is found at $T_{c}^{0} = 0.138\,\text{GeV}$. Also, the scale factor expected to be $z_{0} \in (1,2)$ \cite{Mukherjee:2019eou} and here we simply take $z_{0} = 1.5$.

\begin{figure}
\centering
\includegraphics[width=0.43\textwidth]{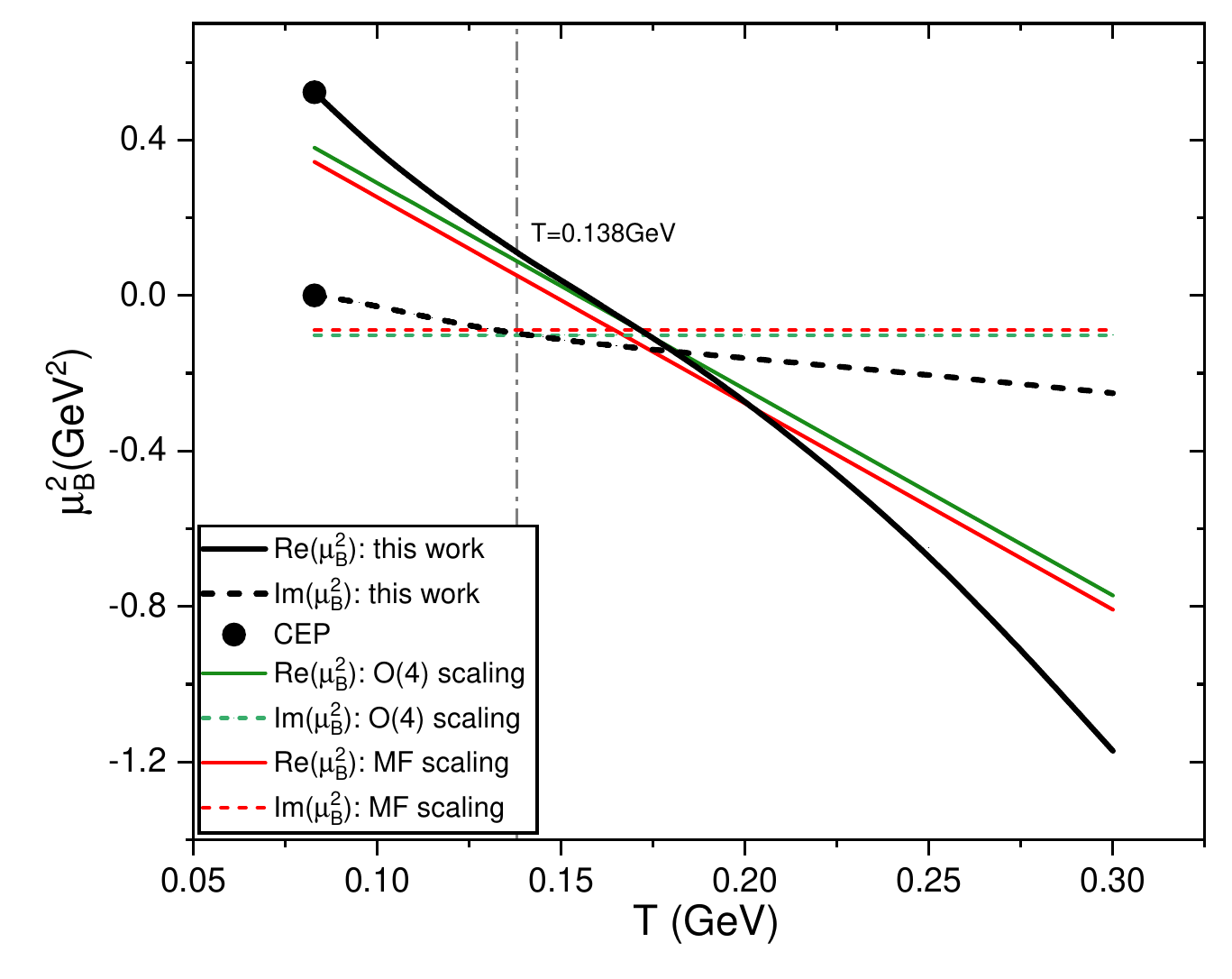}
\vspace*{-4mm}
\caption{The obtained temperature dependence of the squared complex chemical potential of the LYE singularity $\mu_{LYE}^2$, including its real part (solid line) and imaginary part (dashed line), respectively. The dashed vertical line is at $T_c^0=138$ MeV. The case of CEP is also marked with the black solid dots. }
\label{muLYS}
\end{figure}

On the other hand, the quark gap equation under the current truncation scheme is expected to be in the mean field universality class.
To investigate this, we take Eq.~\eqref{O4sca} to explore the scaling behavior of the singularities. In specific, we study the temperature dependence of $\mu_{LYE}^2$ for both its real and the imaginary part, which is shown explicitly in \fig{muLYS}. For the real part of $\mu_{LYE}^2$, the $T$ dependence is approximately linear, as expected in Eq.~\eqref{O4sca}.
%
%
However, the $T$ dependence of $\im \mu_{LYE}^2$ is not negligible, not only close to the CEP but also at  $T = T_{c}^{0} = 0.138\,\textrm{GeV}$.
This suggests an additional temperature dependence of the $z_{c}$ and thus a deviation of the scaling behavior to either the 3D $O(N)$ or the mean field universality class, for a physical quark mass.  \\

\begin{figure}
    \centering
    \includegraphics[width=0.43\textwidth]{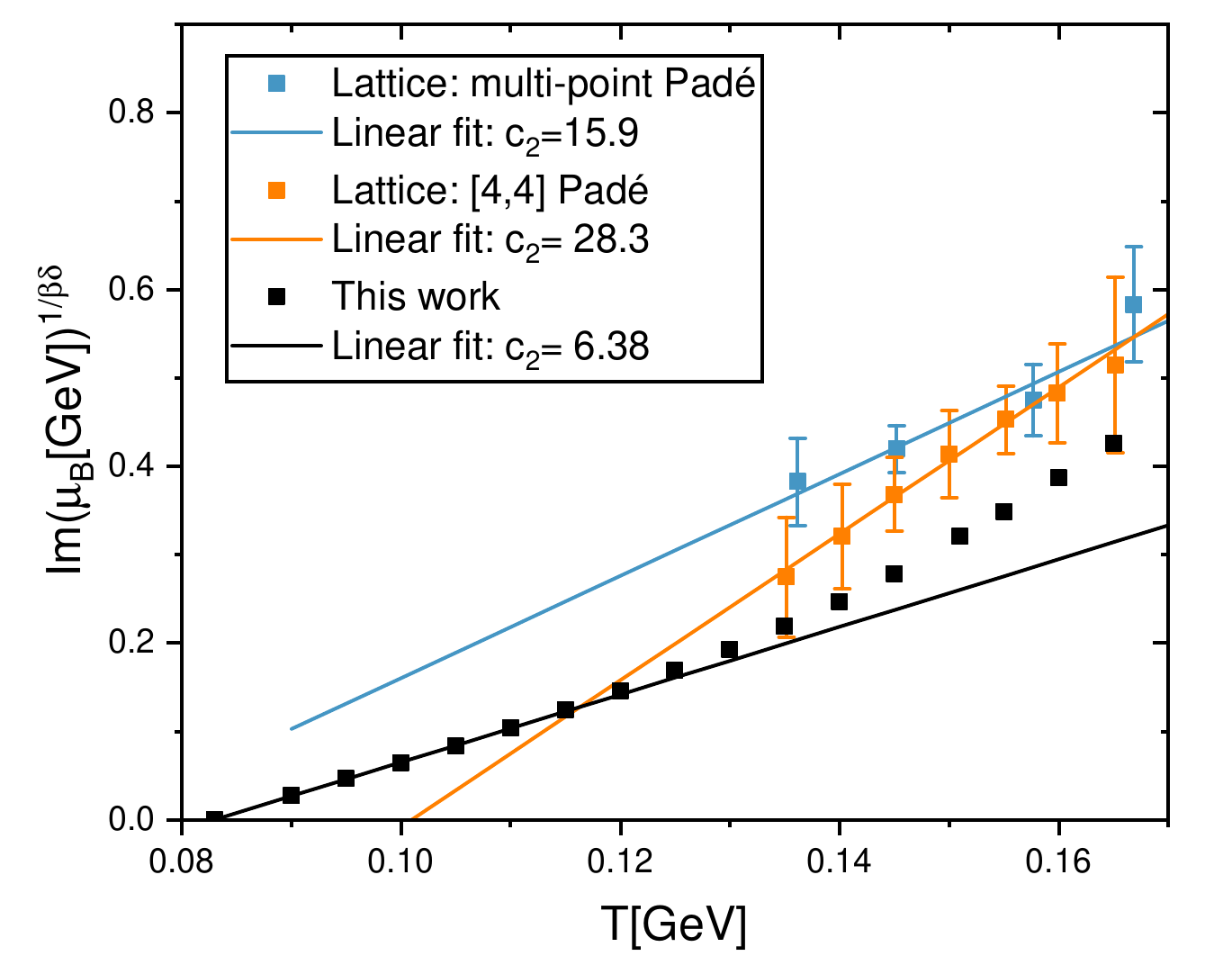}
    \caption{The scaling behavior of the imaginary part of the chemical potential and a linear fit of $\mathrm{Im}[\mu_B]^{1/(\beta\delta)}$ and $T$ together with the LYE singularities from lattice simulations~\cite{Goswami:2024jlc,Dimopoulos:2022,Bollweg:2022rps}.  The parameter  $c_{2}$ can be directly  extracted from the slope, and applied to estimate the CEP location.
    Generally speaking,  for larger $c_2$ , one typically gets a higher temperature for CEP. }
    \label{fig:crireg}
\end{figure}

Finally, consider the region near the CEP, the universality class is expected to be Z(2) symmetry. By performing the linear mapping of the thermodynamic parameters $(T,\mu_B)$ to the scaling variables $(t,h)$:
\begin{equation}
    \left(\begin{array}{c}
t\\[1mm]
h
\end{array}\right)=\mathbb{M}\left(\begin{array}{c}
T - \Tcep \\[1mm]
\mu_{B}^{} - \mu_{B,CEP}^{}
\end{array}\right),\, \quad \mathbb{M}=\left(\begin{array}{ll}
t_{T} & t_{\mu} \\[1mm]
h_{T} & h_{\mu}
\end{array}\right) \, , \quad
\end{equation}
one can derive the Z(2) scaling formula of the LYE singularities\cite{Basar:2021hdf,Stephanov:2006}:
\begin{equation}\label{eq: z2scal}
\begin{aligned}
\mu_{LYE}^{} & \sim \mu_{B,CEP}^{} - c_{1} (T-\Tcep) + \textrm{i} c_{2} (T - \Tcep)^{\beta \delta},\\[1mm]
\text{with } & c_{1} =\frac{h_{T}}{h_{\mu}}, \quad c_{2} = x_{LY}^{} \frac{t_{\mu}{}^{\beta\delta}}{h_{\mu}}\left(\frac{t_{T}}{t_{\mu}}-\frac{h_{T}}{h_{\mu}}\right)^{\beta\delta}.
\end{aligned}
\end{equation}
where $x_{LY}^{}$ is the imaginary part of LYE singularity in Ising model and reads $x_{LY}^{} = |z_{c}|^{-\beta\delta}$.

\begin{figure}
    \centering
    \includegraphics[width=0.43\textwidth]{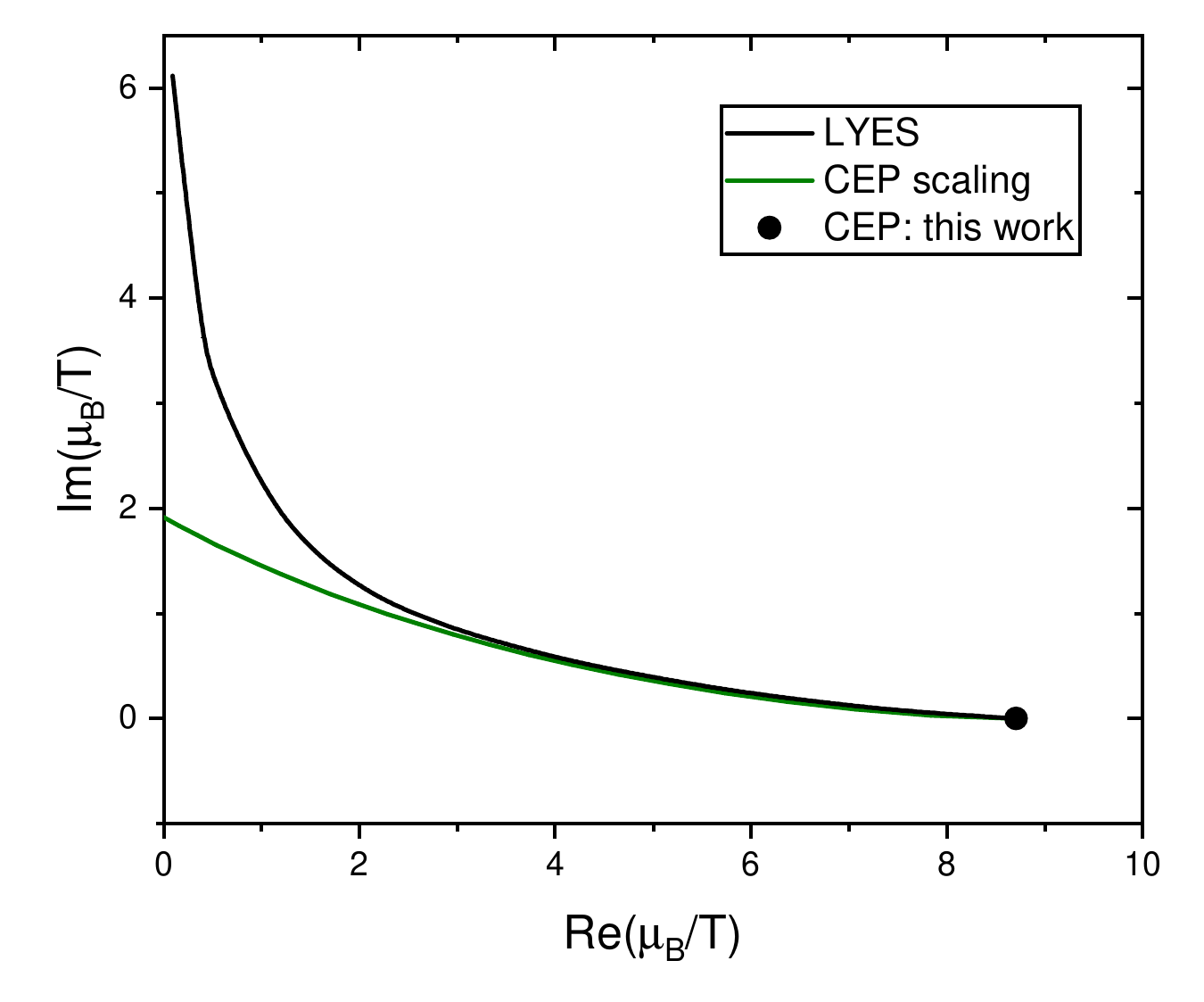}
    \caption{The LYE singularities trajectory (black), which is well compared with the extracted CEP scaling trajectory (green) from \Eq{eq: z2scal}. } 
    \label{fig:z2scal}
\end{figure}

By analysing the scaling behavior of the imaginary part of the chemical potential with respect to the reduced temperature as shown in \fig{fig:crireg}, 
the critical region can be determined as $\ln((T-\Tcep)/\Tcep)<-1$, i.e. $T\in [83,110]\,\text{MeV}$.
Also, we perform a linear fit between the calculated $\ln |\im \mu_{LYE}|$ and $\ln |T-T_{CEP}|$ in the vicinity of CEP to extract the critical exponents $\beta\delta$.
This yields $\beta\delta=1.38$, which slightly  differs from the exact value of Z(2) universality class shown in \tab{uni_para} due to the lack of detailed critical information in current truncation scheme. 
However, the CEP scaling behavior is clear and  the constants $c_1$ and $c_2$ in \Eq{eq: z2scal} can be estimated  by fitting with our results within the critical region. 
The corresponding CEP scaling trajectory is obtained and extrapolated towards small $\re\,\mu_{B}^{}$,  shown as the green line in \fig{fig:z2scal}, 
which shows that the CEP scaling extrapolation is consistent with the calculated trajectory near CEP.
%
%

%
%

Our results show that the Z(2) scaling works well for the LYE singularities, and therefore, one may directly apply the Ising parameterization to study the CEP.  
Firstly, we apply the following parameterization as it can well describe the equation of state of QCD~\cite{Lu:2023msn,Parotto:2020,Rehr:1973zz}. Typically, we take the form as shown in Ref.~\cite{Lu:2023msn}:
\begin{equation}\label{eq:Isingmap}
  \begin{aligned}
  & \frac{\,\mu_{B}^{} - \mu_{B,CEP}^{}\,}{\mu_{B,CEP}^{}} = - t \omega \rho \cos{\alpha_{1}} - h \omega \cos{\alpha_{2}}, \\
  & \frac{T - \Tcep}{\Tcep} =  f_{\mathrm{PT}}^{} (t) + h \omega \sin{\alpha_{2}},
\end{aligned}
\end{equation}
with $\omega=1$, $\rho=2$,  $\alpha_{1} = \tan^{-1} ( 2\kappa \mu_{B,CEP}^{} /T_{c}(0) )  $ and $\alpha_{2} = \alpha_{1} + \pi/2$ in our case.
The mapping function $f_{\mathrm{PT}}^{}$ is calibrated so that at $h=0$, Eqs.~(\ref{eq:Isingmap}) becomes exactly the parametric equations of the phase transition line in Eq.~(\ref{eq:PTline}), which requires
\begin{equation}\label{eq:nlmap}
%
    f_{\mathrm{PT}}^{}(t) = \frac{\,\mu_{B,CEP}^{}\,}{2 \Tcep} \left( 2 - t \omega \rho \cos {\alpha_{1}} \right) \, t \omega \rho \sin {\alpha_{1}} .
\end{equation}
This gives then immediately the dimensionless slope parameter $\overline{c\,}_{2}$ as:
%
%
%
\begin{equation}
  \begin{aligned}
\overline{c\,}_{2} & = x_{LY}^{} \frac{ (\frac{T_{CEP}}{\,\mu_{B,CEP}^{}\,})^{\beta\delta} \omega \sin{\alpha_{1}} }{ (\omega \rho  \sin{\alpha_1})^{\beta\delta} } \left( \frac{T_{CEP}}{\mu_{B,CEP}^{}} \cot^{2}{\alpha_{1}} + 1  \right),
   \end{aligned}
\end{equation}
with $\overline{c\,}_{2} = c_{2} \frac{(T_{CEP})^{\beta\delta}}{\mu_{B,CEP}^{}}$. We note that this dimensionless treatment allows us to estimate the CEP location which is independent of the scale setting in the truncation scheme.
Together with the parameterization of phase transition line, we have
\begin{equation}
T_{CEP} = T_{c} (0) - \kappa\frac{ \mu_{B,CEP}^{\,2} }{T_{c}(0)},
\end{equation}
the location of the CEP can be then determined after one extracts the $c_{2}$ from the numerical results of LYE singularities in QCD. 
In \fig{fig:crireg}, we show the extraction of the $c_{2}$ from the LYE singularities.

Through the linear fit between $\im[\mu_{B}^{}]^{1/(\beta\delta)}$ and $T$, one can extract the $c_{2}$ directly from the slope,
and we get $c_{2} =6.38\, \textrm{GeV}^{1-\beta\delta}$ and $\overline{c\,}_{2} = 0.28$.
Now if here we adopt  $\beta\delta$ and $z_{c}$ from Z(2) universality class and  the phase transition line curvature $\kappa=0.016$ and $T_{c}(0)=156 \, \textrm{MeV}$,  one gets the CEP at:
\begin{eqnarray}
(T_{CEP}, \mu_{B,CEP}^{}) = (118,606) \,\,\mathrm{MeV} \, .
\end{eqnarray}
This indicates that with the $ \overline{c}_{2}$ extracted in our work together with the up to date knowledge of the phase transition line and the correct scaling behavior,
the obtained CEP location is in precise agreement with the up to date results from the functional QCD approaches~\cite{Gao:2020fbl,Gunkel:2021oya,Gao:2020qsj,Fu:2019hdw}.
Besides, the rescaled $c^{\prime}_{2}$ based on the estimated CEP  now becomes $c^{\prime}_{2} = 5.0 \,\textrm{GeV}^{1-\beta\delta}$.
It needs also to mention that one may get a much larger $c_{2}$ with the current results of LYE singularities from lattice QCD simulations~\cite{Goswami:2024jlc,Dimopoulos:2022,Bollweg:2022rps}.
It is mainly because   the current  lattice QCD results  are obtained in  too high temperature  region and has not reached the scaling region as shown in Fig.~\ref{fig:crireg}.
For the LYE singularities next to the CEP scaling region, the slope is steeper which stands for a large value of $c_{2}$.

\section{Analysis on the parameter dependence in the truncation}\label{sec:modelparam}

In this section, we explore the effects of the parameters in the Ansatz of the quark-gluon interaction vertex in~\Eq{vertex} on the properties of the LYE singularities.

In \Eq{vertex}, we note that the parameters $\beta$, $\delta$ and $\Lambda$ are tightly constrained from the running of the quark-gluon vertex function $\Gamma(k^2)$ in \eq{vertex} at large momentum scale.
Specifically, $\beta$ and $\delta$ the anomalous dimensions of the vertex, along with $\Lambda$ which is related to the QCD scale parameter.
The remaining parameters $d_{1}$ and $d_{2}$ are phenomenological.
In previous sections, their values are taken as the parameter set for (2+1)-flavour QCD, which can well describe the meson spectrum. This parameter set will be denoted as set (I) in the following discussion.
We then move on to investigate the effect of changing $d_1$ and $d_2$ on the results, with the cases of $(d_1,d_2)$ considered in \fig{fig:vtx}.
The obtained variation behavior of some of the characteristic quantities with respect to the $d_{1}$ and $d_{2}$ are listed in Table~\ref{tab:dres}.

\begin{table*}[t!]
  \begin{center}
    \caption{Analysis on the parameter dependence $d_1$ and $d_2$ for the chiral phase transition, and the properties of the Lee-Yang edge singularities.}
    \vspace*{-2mm}
    \begin{tabular}{|c|c|c|c|c|c|c|c|c|}
      \hline
      set & $(d_1,d_2)$ GeV$^2$ & $M_q(0)$ GeV& $T_c(0)$ MeV & $(T_{CEP},\mu_{B,CEP})$ MeV&$\beta\delta$&$c_2\, \textrm{GeV}^{1-\beta\delta}$&$\overline{c\,}_2$&$(T'_{CEP},\mu'_{B,CEP})$ MeV\\
      \hline
      (I) & (6.8,0.5) & 0.36 & 155.7 & (83,723)&1.38&6.38&0.28&(118,606)\\
        \hline
      (II) & (7.2,0.57) & 0.36 & 155.8 &(83,741) &1.36 & 6.41 & 0.30 &(119,602)\\
     \hline
     \hline
      (III) & (7.2,0.5) & 0.40 & 168 &(99,708) &1.38 &6.8 & 0.39 &(122,578)\\
      \hline
      (IV) & (8.0,0.5) & 0.47 & 195 & (130,642)&1.36&7.1&0.70&(139,540)\\
      \hline
    \end{tabular}
  \label{tab:dres}
  \end{center}

\end{table*}

Firstly, we consider the case of changing $d_1$ togther with $d_2$, while keeping the obtained vacuum quark mass at zero momentum $M_q(0)$ unchanged (II).
With such a setup, we found that the obtained chiral phase transition temperature at zero chemical potential $T_c(0)$ and the location of CEP are very close to the ones obtained from the set (I), see in \tab{tab:dres}.
Therefore, the sets (I) and (II) provide explicitly an error estimate on the infrared behaviour of the vertex in (2+1)-flavour QCD.
Secondly, we found that the $d_{2}$ affects the results less significantly than the $d_{1}$, since the $d_{2}$ stands for the running behaviour of the interaction vertex, while $d_{1}$ reflects on the vertex strength directly.
Therefore, we investigate several $d_{1}$ values while keeping $d_{2}$ unchanged to test the robustness of the CEP scaling, which is referred to as sets (III) and (IV) in \tab{tab:dres}.

Comparing the results for sets (I)-(IV), it can be found that different $d_{1}$ and $d_{2}$ parameters yield almost the same critical exponents:
\begin{equation}
    \beta \delta \approx 1.37\pm0.01,
\end{equation}
with less than 2\% error, implying that the observed CEP scaling of the LYE singularities is a robust property.
Moreover, it is clear by comparing the results in the sets (I) and (II) that with the same quark mass $M_{q}(0)$ obtained, the chiral phase transition line as well as the location of the LYE singularities and their scaling parameters $c_2$ and $\bar{c}_2$ are independent to the choice of parameters.
We also show that the location of the extracted CEP according to the Z(2) universality class (see the above \Sec{anlys}) is also unchanged from set (I) to set (II).

On the other hand, by changing the $d_{1}$ with the $d_{2}$ being maintained constant,
the parameter set deviates from the ones for the (2+1)-flavour QCD, which is demonstrated in the change of the chiral phase transition line, i.e. $T_{c}(0)$ and the CEP location.
The scaling parameters $c_{2}$ and $\overline{c\,}_2$ will also change: generally speaking, they become larger as $d_{1}$ increases, or as $d_{2}$ decreases.    
We emphasise, however, that the CEP scaling is still observed, with a similar trend of the LYE singularities at small $\im{\mu_B}$. The CEP scaling found here is robust against the choices of parameters.

\section{Summary}\label{sum}

We calculate the quark gap equation at complex chemical potential with physical quark mass via the Dyson-Schwinger equation approach of QCD.
We firstly check the analytic continuity of the phase transition line from imaginary to real chemical potential, which is required in the lattice QCD simulations.
Our result shows that the curvature of the chiral phase transition line extracted from the imaginary chemical potential is consistent with that from the real chemical potential.

Moreover, we study the Lee-Yang edge singularities which have been shown to be related to the phase structure at the real chemical potential.
We compute the location of the Lee-Yang edge singularities in the complex $\mu_{B}^{}$ plane at different temperatures.
Our result manifests that, at the temperature where there is a crossover at the real chemical potential, the Lee-Yang edge singularity is located at the complex plane of the chemical potential with an imaginary part.
The absolute value of the imaginary part becomes smaller as the temperature decreases, and finally vanishes at the critical end point temperature.

We then take the Pad\'{e} polynomials to see whether it is possible to extrapolate the CEP location with the limited range of the Lee-Yang edge singularities.
Our result indicates that additional information at high $\mu_{B}^{}$, especially beyond the range of current best lattice QCD expansions, is required from the theoretical approaches for a precise determination of the CEP.

We further check the scaling behavior of the trajectory of Lee-Yang edge singularities.
We compute the radius of the convergence and find that it changes for different temperatures.
The value of the convergence radius is also found to be inconsistent with any scaling value for either 3D $O(N)$ or mean field universality class.
However, it is found that the LYE singularities follows  the CEP scaling as the Z(2) universality class.
Such a scaling behaviour is further confirmed by changing the phenomenological parameters $d_1$ and $d_2$ in the truncation scheme.
Therefore, one may take the Ising model to study the CEP of QCD.
Here we implement a mapping of the QCD order parameters with Ising parameterization, and shows that after extracting $c_{2}$ from the current data of the LYE singularities location together with the curvature of the phase transition line,
one may extract the location of the CEP with the information of Ising universality class.

For a high temperature, the real part of the Lee-Yang edge singularities becomes smaller and the imaginary part increases which is consistent with the perturbative analysis.
The imaginary part is expected to be saturated at $\im\,\mu_{B}^{} =3 \pi T$ which requires a careful scan for a wide range of the complex chemical potential.
This may also reveal the periodicity  of the thermodynamic quantities which represents the Z($N_{c}$) symmetry and potentially the deconfinement.
The related investigation is under progress.

\medskip

\section{Acknowledgements}

YL and FG thank the other members of the  fQCD collaboration~\cite{fQCD} for fruitful discussions.
This work  is supported by the National Natural Science Foundation of China under Grants  No. 12247107, No. 12175007. FG is also supported by the National  Science Foundation of China under Grants  No. 12305134.

\bibliographystyle{sn-mathphys-num}
\bibliography{Rev2-EPJC-24-02-039}


\begin{thebibliography}{52}
\ifx \bisbn   \undefined \def \bisbn  #1{ISBN #1}\fi
\ifx \binits  \undefined \def \binits#1{#1}\fi
\ifx \bauthor  \undefined \def \bauthor#1{#1}\fi
\ifx \batitle  \undefined \def \batitle#1{#1}\fi
\ifx \bjtitle  \undefined \def \bjtitle#1{#1}\fi
\ifx \bvolume  \undefined \def \bvolume#1{\textbf{#1}}\fi
\ifx \byear  \undefined \def \byear#1{#1}\fi
\ifx \bissue  \undefined \def \bissue#1{#1}\fi
\ifx \bfpage  \undefined \def \bfpage#1{#1}\fi
\ifx \blpage  \undefined \def \blpage #1{#1}\fi
\ifx \burl  \undefined \def \burl#1{\textsf{#1}}\fi
\ifx \doiurl  \undefined \def \doiurl#1{\url{https://doi.org/#1}}\fi
\ifx \betal  \undefined \def \betal{\textit{et al.}}\fi
\ifx \binstitute  \undefined \def \binstitute#1{#1}\fi
\ifx \binstitutionaled  \undefined \def \binstitutionaled#1{#1}\fi
\ifx \bctitle  \undefined \def \bctitle#1{#1}\fi
\ifx \beditor  \undefined \def \beditor#1{#1}\fi
\ifx \bpublisher  \undefined \def \bpublisher#1{#1}\fi
\ifx \bbtitle  \undefined \def \bbtitle#1{#1}\fi
\ifx \bedition  \undefined \def \bedition#1{#1}\fi
\ifx \bseriesno  \undefined \def \bseriesno#1{#1}\fi
\ifx \blocation  \undefined \def \blocation#1{#1}\fi
\ifx \bsertitle  \undefined \def \bsertitle#1{#1}\fi
\ifx \bsnm \undefined \def \bsnm#1{#1}\fi
\ifx \bsuffix \undefined \def \bsuffix#1{#1}\fi
\ifx \bparticle \undefined \def \bparticle#1{#1}\fi
\ifx \barticle \undefined \def \barticle#1{#1}\fi
\bibcommenthead
\ifx \bconfdate \undefined \def \bconfdate #1{#1}\fi
\ifx \botherref \undefined \def \botherref #1{#1}\fi
\ifx \url \undefined \def \url#1{\textsf{#1}}\fi
\ifx \bchapter \undefined \def \bchapter#1{#1}\fi
\ifx \bbook \undefined \def \bbook#1{#1}\fi
\ifx \bcomment \undefined \def \bcomment#1{#1}\fi
\ifx \oauthor \undefined \def \oauthor#1{#1}\fi
\ifx \citeauthoryear \undefined \def \citeauthoryear#1{#1}\fi
\ifx \endbibitem  \undefined \def \endbibitem {}\fi
\ifx \bconflocation  \undefined \def \bconflocation#1{#1}\fi
\ifx \arxivurl  \undefined \def \arxivurl#1{\textsf{#1}}\fi
\csname PreBibitemsHook\endcsname

\bibitem[\protect\citeauthoryear{Aarts et~al.}{2023}]{Aarts:2023vsf}
\begin{barticle}
\bauthor{\bsnm{Aarts}, \binits{G.}}, \betal:
\batitle{{Phase Transitions in Particle Physics}: {Results and Perspectives
  from Lattice Quantum Chromo-Dynamics}}.
\bjtitle{Prog. Part. Nucl. Phys.}
\bvolume{133},
\bfpage{104070}
(\byear{2023})
\doiurl{10.1016/j.ppnp.2023.104070}
{\href{https://arxiv.org/abs/2301.04382}{{arXiv:2301.04382}}}
{[hep-lat]}
\end{barticle}
\endbibitem

\bibitem[\protect\citeauthoryear{Hippert et~al.}{2023}]{Hippert:2023bel}
\begin{botherref}
\oauthor{\bsnm{Hippert}, \binits{M.}},
\oauthor{\bsnm{Grefa}, \binits{J.}},
\oauthor{\bsnm{Manning}, \binits{T.A.}},
\oauthor{\bsnm{Noronha}, \binits{J.}},
\oauthor{\bsnm{Noronha-Hostler}, \binits{J.}},
\oauthor{\bsnm{Portillo~Vazquez}, \binits{I.}},
\oauthor{\bsnm{Ratti}, \binits{C.}},
\oauthor{\bsnm{Rougemont}, \binits{R.}},
\oauthor{\bsnm{Trujillo}, \binits{M.}}:
{Bayesian location of the QCD critical point from a holographic perspective}
(2023)
{\href{https://arxiv.org/abs/2309.00579}{{arXiv:2309.00579}}}
{[nucl-th]}
\end{botherref}
\endbibitem

\bibitem[\protect\citeauthoryear{Huang and Zhuang}{2023}]{Huang:2023ogw}
\begin{barticle}
\bauthor{\bsnm{Huang}, \binits{M.}},
\bauthor{\bsnm{Zhuang}, \binits{P.}}:
\batitle{{QCD Matter and Phase Transitions under Extreme Conditions}}.
\bjtitle{Symmetry}
\bvolume{15}(\bissue{2}),
\bfpage{541}
(\byear{2023})
\doiurl{10.3390/sym15020541}
\end{barticle}
\endbibitem

\bibitem[\protect\citeauthoryear{Fu}{2022}]{Fu:2022gou}
\begin{barticle}
\bauthor{\bsnm{Fu}, \binits{W.-j.}}:
\batitle{{QCD at finite temperature and density within the fRG approach: an
  overview}}.
\bjtitle{Commun. Theor. Phys.}
\bvolume{74}(\bissue{9}),
\bfpage{097304}
(\byear{2022})
\doiurl{10.1088/1572-9494/ac86be}
{\href{https://arxiv.org/abs/2205.00468}{{arXiv:2205.00468}}}
{[hep-ph]}
\end{barticle}
\endbibitem

\bibitem[\protect\citeauthoryear{Lovato et~al.}{2022}]{Lovato:2022vgq}
\begin{botherref}
\oauthor{\bsnm{Lovato}, \binits{A.}}, et al.:
{Long Range Plan: Dense matter theory for heavy-ion collisions and neutron
  stars}
(2022)
{\href{https://arxiv.org/abs/2211.02224}{{arXiv:2211.02224}}}
{[nucl-th]}
\end{botherref}
\endbibitem

\bibitem[\protect\citeauthoryear{Adamczewski-Musch
  et~al.}{2019}]{HADES:2019auv}
\begin{barticle}
\bauthor{\bsnm{Adamczewski-Musch}, \binits{J.}}, \betal:
\batitle{{Probing dense baryon-rich matter with virtual photons}}.
\bjtitle{Nature Phys.}
\bvolume{15}(\bissue{10}),
\bfpage{1040}--\blpage{1045}
(\byear{2019})
\doiurl{10.1038/s41567-019-0583-8}
\end{barticle}
\endbibitem

\bibitem[\protect\citeauthoryear{Fischer}{2019}]{Fischer:2018sdj}
\begin{barticle}
\bauthor{\bsnm{Fischer}, \binits{C.S.}}:
\batitle{{QCD at finite temperature and chemical potential from
  Dyson\textendash{}Schwinger equations}}.
\bjtitle{Prog. Part. Nucl. Phys.}
\bvolume{105},
\bfpage{1}--\blpage{60}
(\byear{2019})
\doiurl{10.1016/j.ppnp.2019.01.002}
{\href{https://arxiv.org/abs/1810.12938}{{arXiv:1810.12938}}}
{[hep-ph]}
\end{barticle}
\endbibitem

\bibitem[\protect\citeauthoryear{Luo and Xu}{2017}]{Luo:2017faz}
\begin{barticle}
\bauthor{\bsnm{Luo}, \binits{X.}},
\bauthor{\bsnm{Xu}, \binits{N.}}:
\batitle{{Search for the QCD Critical Point with Fluctuations of Conserved
  Quantities in Relativistic Heavy-Ion Collisions at RHIC : An Overview}}.
\bjtitle{Nucl. Sci. Tech.}
\bvolume{28}(\bissue{8}),
\bfpage{112}
(\byear{2017})
\doiurl{10.1007/s41365-017-0257-0}
{\href{https://arxiv.org/abs/1701.02105}{{arXiv:1701.02105}}}
{[nucl-ex]}
\end{barticle}
\endbibitem

\bibitem[\protect\citeauthoryear{Braun-Munzinger
  et~al.}{2016}]{Braun-Munzinger:2015hba}
\begin{barticle}
\bauthor{\bsnm{Braun-Munzinger}, \binits{P.}},
\bauthor{\bsnm{Koch}, \binits{V.}},
\bauthor{\bsnm{Sch\"afer}, \binits{T.}},
\bauthor{\bsnm{Stachel}, \binits{J.}}:
\batitle{{Properties of hot and dense matter from relativistic heavy ion
  collisions}}.
\bjtitle{Phys. Rept.}
\bvolume{621},
\bfpage{76}--\blpage{126}
(\byear{2016})
\doiurl{10.1016/j.physrep.2015.12.003}
{\href{https://arxiv.org/abs/1510.00442}{{arXiv:1510.00442}}}
{[nucl-th]}
\end{barticle}
\endbibitem

\bibitem[\protect\citeauthoryear{Schaefer and Wagner}{2009}]{Schaefer:2008ax}
\begin{barticle}
\bauthor{\bsnm{Schaefer}, \binits{B.-J.}},
\bauthor{\bsnm{Wagner}, \binits{M.}}:
\batitle{{On the QCD phase structure from effective models}}.
\bjtitle{Prog. Part. Nucl. Phys.}
\bvolume{62},
\bfpage{381}
(\byear{2009})
\doiurl{10.1016/j.ppnp.2008.12.009}
{\href{https://arxiv.org/abs/0812.2855}{{arXiv:0812.2855}}}
{[hep-ph]}
\end{barticle}
\endbibitem

\bibitem[\protect\citeauthoryear{Borsanyi et~al.}{2020}]{Borsanyi:2020fev}
\begin{barticle}
\bauthor{\bsnm{Borsanyi}, \binits{S.}},
\bauthor{\bsnm{Fodor}, \binits{Z.}},
\bauthor{\bsnm{Guenther}, \binits{J.N.}},
\bauthor{\bsnm{Kara}, \binits{R.}},
\bauthor{\bsnm{Katz}, \binits{S.D.}},
\bauthor{\bsnm{Parotto}, \binits{P.}},
\bauthor{\bsnm{Pasztor}, \binits{A.}},
\bauthor{\bsnm{Ratti}, \binits{C.}},
\bauthor{\bsnm{Szabo}, \binits{K.K.}}:
\batitle{{QCD Crossover at Finite Chemical Potential from Lattice
  Simulations}}.
\bjtitle{Phys. Rev. Lett.}
\bvolume{125}(\bissue{5}),
\bfpage{052001}
(\byear{2020})
\doiurl{10.1103/PhysRevLett.125.052001}
{\href{https://arxiv.org/abs/2002.02821}{{arXiv:2002.02821}}}
{[hep-lat]}
\end{barticle}
\endbibitem

\bibitem[\protect\citeauthoryear{Bazavov et~al.}{2019}]{HotQCD:2018pds}
\begin{barticle}
\bauthor{\bsnm{Bazavov}, \binits{A.}}, \betal:
\batitle{{Chiral crossover in QCD at zero and non-zero chemical potentials}}.
\bjtitle{Phys. Lett. B}
\bvolume{795},
\bfpage{15}--\blpage{21}
(\byear{2019})
\doiurl{10.1016/j.physletb.2019.05.013}
{\href{https://arxiv.org/abs/1812.08235}{{arXiv:1812.08235}}}
{[hep-lat]}
\end{barticle}
\endbibitem

\bibitem[\protect\citeauthoryear{Gao and Pawlowski}{2021}]{Gao:2020fbl}
\begin{barticle}
\bauthor{\bsnm{Gao}, \binits{F.}},
\bauthor{\bsnm{Pawlowski}, \binits{J.M.}}:
\batitle{{Chiral phase structure and critical end point in QCD}}.
\bjtitle{Phys. Lett. B}
\bvolume{820},
\bfpage{136584}
(\byear{2021})
\doiurl{10.1016/j.physletb.2021.136584}
{\href{https://arxiv.org/abs/2010.13705}{{arXiv:2010.13705}}}
{[hep-ph]}
\end{barticle}
\endbibitem

\bibitem[\protect\citeauthoryear{Gunkel and Fischer}{2021}]{Gunkel:2021oya}
\begin{barticle}
\bauthor{\bsnm{Gunkel}, \binits{P.J.}},
\bauthor{\bsnm{Fischer}, \binits{C.S.}}:
\batitle{{Locating the critical endpoint of QCD: Mesonic backcoupling
  effects}}.
\bjtitle{Phys. Rev. D}
\bvolume{104}(\bissue{5}),
\bfpage{054022}
(\byear{2021})
\doiurl{10.1103/PhysRevD.104.054022}
{\href{https://arxiv.org/abs/2106.08356}{{arXiv:2106.08356}}}
{[hep-ph]}
\end{barticle}
\endbibitem

\bibitem[\protect\citeauthoryear{Gao and Pawlowski}{2020}]{Gao:2020qsj}
\begin{barticle}
\bauthor{\bsnm{Gao}, \binits{F.}},
\bauthor{\bsnm{Pawlowski}, \binits{J.M.}}:
\batitle{{QCD phase structure from functional methods}}.
\bjtitle{Phys. Rev. D}
\bvolume{102}(\bissue{3}),
\bfpage{034027}
(\byear{2020})
\doiurl{10.1103/PhysRevD.102.034027}
{\href{https://arxiv.org/abs/2002.07500}{{arXiv:2002.07500}}}
{[hep-ph]}
\end{barticle}
\endbibitem

\bibitem[\protect\citeauthoryear{Fu et~al.}{2020}]{Fu:2019hdw}
\begin{barticle}
\bauthor{\bsnm{Fu}, \binits{W.-j.}},
\bauthor{\bsnm{Pawlowski}, \binits{J.M.}},
\bauthor{\bsnm{Rennecke}, \binits{F.}}:
\batitle{{QCD phase structure at finite temperature and density}}.
\bjtitle{Phys. Rev. D}
\bvolume{101}(\bissue{5}),
\bfpage{054032}
(\byear{2020})
\doiurl{10.1103/PhysRevD.101.054032}
{\href{https://arxiv.org/abs/1909.02991}{{arXiv:1909.02991}}}
{[hep-ph]}
\end{barticle}
\endbibitem

\bibitem[\protect\citeauthoryear{Cai et~al.}{2022}]{Cai:2022omk}
\begin{barticle}
\bauthor{\bsnm{Cai}, \binits{R.-G.}},
\bauthor{\bsnm{He}, \binits{S.}},
\bauthor{\bsnm{Li}, \binits{L.}},
\bauthor{\bsnm{Wang}, \binits{Y.-X.}}:
\batitle{{Probing QCD critical point and induced gravitational wave by black
  hole physics}}.
\bjtitle{Phys. Rev. D}
\bvolume{106}(\bissue{12}),
\bfpage{121902}
(\byear{2022})
\doiurl{10.1103/PhysRevD.106.L121902}
{\href{https://arxiv.org/abs/2201.02004}{{arXiv:2201.02004}}}
{[hep-th]}
\end{barticle}
\endbibitem

\bibitem[\protect\citeauthoryear{Yang and Lee}{1952}]{Yang:1952be}
\begin{barticle}
\bauthor{\bsnm{Yang}, \binits{C.-N.}},
\bauthor{\bsnm{Lee}, \binits{T.D.}}:
\batitle{{Statistical theory of equations of state and phase transitions. 1.
  Theory of condensation}}.
\bjtitle{Phys. Rev.}
\bvolume{87},
\bfpage{404}--\blpage{409}
(\byear{1952})
\doiurl{10.1103/PhysRev.87.404}
\end{barticle}
\endbibitem

\bibitem[\protect\citeauthoryear{Lee and Yang}{1952}]{Lee:1952ig}
\begin{barticle}
\bauthor{\bsnm{Lee}, \binits{T.D.}},
\bauthor{\bsnm{Yang}, \binits{C.-N.}}:
\batitle{{Statistical theory of equations of state and phase transitions. 2.
  Lattice gas and Ising model}}.
\bjtitle{Phys. Rev.}
\bvolume{87},
\bfpage{410}--\blpage{419}
(\byear{1952})
\doiurl{10.1103/PhysRev.87.410}
\end{barticle}
\endbibitem

\bibitem[\protect\citeauthoryear{Clarke et~al.}{2023}]{Clarke:2023noy}
\begin{barticle}
\bauthor{\bsnm{Clarke}, \binits{D.A.}},
\bauthor{\bsnm{Zambello}, \binits{K.}},
\bauthor{\bsnm{Dimopoulos}, \binits{P.}},
\bauthor{\bsnm{Di~Renzo}, \binits{F.}},
\bauthor{\bsnm{Goswami}, \binits{J.}},
\bauthor{\bsnm{Nicotra}, \binits{G.}},
\bauthor{\bsnm{Schmidt}, \binits{C.}},
\bauthor{\bsnm{Singh}, \binits{S.}}:
\batitle{{Determination of Lee-Yang edge singularities in QCD by rational
  approximations}}.
\bjtitle{PoS}
\bvolume{LATTICE2022},
\bfpage{164}
(\byear{2023})
\doiurl{10.22323/1.430.0164}
{\href{https://arxiv.org/abs/2301.03952}{{arXiv:2301.03952}}}
{[hep-lat]}
\end{barticle}
\endbibitem

\bibitem[\protect\citeauthoryear{Schmidt et~al.}{2023}]{Schmidt:2022ogw}
\begin{barticle}
\bauthor{\bsnm{Schmidt}, \binits{C.}},
\bauthor{\bsnm{Clarke}, \binits{D.A.}},
\bauthor{\bsnm{Nicotra}, \binits{G.}},
\bauthor{\bsnm{Di~Renzo}, \binits{F.}},
\bauthor{\bsnm{Dimopoulos}, \binits{P.}},
\bauthor{\bsnm{Singh}, \binits{S.}},
\bauthor{\bsnm{Goswami}, \binits{J.}},
\bauthor{\bsnm{Zambello}, \binits{K.}}:
\batitle{{Detecting Critical Points from the Lee\textendash{}Yang Edge
  Singularities in Lattice QCD}}.
\bjtitle{Acta Phys. Polon. Supp.}
\bvolume{16}(\bissue{1}),
\bfpage{1}--\blpage{52}
(\byear{2023})
\doiurl{10.5506/APhysPolBSupp.16.1-A52}
{\href{https://arxiv.org/abs/2209.04345}{{arXiv:2209.04345}}}
{[hep-lat]}
\end{barticle}
\endbibitem

\bibitem[\protect\citeauthoryear{Singh et~al.}{2022}]{Singh:2021pog}
\begin{barticle}
\bauthor{\bsnm{Singh}, \binits{S.}},
\bauthor{\bsnm{Dimopoulos}, \binits{P.}},
\bauthor{\bsnm{Dini}, \binits{L.}},
\bauthor{\bsnm{Di~Renzo}, \binits{F.}},
\bauthor{\bsnm{Goswami}, \binits{J.}},
\bauthor{\bsnm{Nicotra}, \binits{G.}},
\bauthor{\bsnm{Schmidt}, \binits{C.}},
\bauthor{\bsnm{Zambello}, \binits{K.}},
\bauthor{\bsnm{Ziesche}, \binits{F.}}:
\batitle{{Lee-Yang edge singularities in lattice QCD : A systematic study of
  singularities in the complex muB plane using rational approximations.}}
\bjtitle{PoS}
\bvolume{LATTICE2021},
\bfpage{544}
(\byear{2022})
\doiurl{10.22323/1.396.0544}
{\href{https://arxiv.org/abs/2111.06241}{{arXiv:2111.06241}}}
{[hep-lat]}
\end{barticle}
\endbibitem

\bibitem[\protect\citeauthoryear{Roberge and Weiss}{1986}]{Roberge:1986mm}
\begin{barticle}
\bauthor{\bsnm{Roberge}, \binits{A.}},
\bauthor{\bsnm{Weiss}, \binits{N.}}:
\batitle{{Gauge Theories With Imaginary Chemical Potential and the Phases of
  {QCD}}}.
\bjtitle{Nucl. Phys. B}
\bvolume{275},
\bfpage{734}--\blpage{745}
(\byear{1986})
\doiurl{10.1016/0550-3213(86)90582-1}
\end{barticle}
\endbibitem

\bibitem[\protect\citeauthoryear{Fischer et~al.}{2015}]{Fischer:2014vxa}
\begin{barticle}
\bauthor{\bsnm{Fischer}, \binits{C.S.}},
\bauthor{\bsnm{Luecker}, \binits{J.}},
\bauthor{\bsnm{Pawlowski}, \binits{J.M.}}:
\batitle{{Phase structure of QCD for heavy quarks}}.
\bjtitle{Phys. Rev. D}
\bvolume{91}(\bissue{1}),
\bfpage{014024}
(\byear{2015})
\doiurl{10.1103/PhysRevD.91.014024}
{\href{https://arxiv.org/abs/1409.8462}{{arXiv:1409.8462}}}
{[hep-ph]}
\end{barticle}
\endbibitem

\bibitem[\protect\citeauthoryear{Fischer}{2009}]{Fischer:2009wc}
\begin{barticle}
\bauthor{\bsnm{Fischer}, \binits{C.S.}}:
\batitle{{Deconfinement phase transition and the quark condensate}}.
\bjtitle{Phys. Rev. Lett.}
\bvolume{103},
\bfpage{052003}
(\byear{2009})
\doiurl{10.1103/PhysRevLett.103.052003}
{\href{https://arxiv.org/abs/0904.2700}{{arXiv:0904.2700}}}
{[hep-ph]}
\end{barticle}
\endbibitem

\bibitem[\protect\citeauthoryear{Fischer et~al.}{2011}]{Fischer:2011mz}
\begin{barticle}
\bauthor{\bsnm{Fischer}, \binits{C.S.}},
\bauthor{\bsnm{Luecker}, \binits{J.}},
\bauthor{\bsnm{Mueller}, \binits{J.A.}}:
\batitle{{Chiral and deconfinement phase transitions of two-flavour QCD at
  finite temperature and chemical potential}}.
\bjtitle{Phys. Lett. B}
\bvolume{702},
\bfpage{438}--\blpage{441}
(\byear{2011})
\doiurl{10.1016/j.physletb.2011.07.039}
{\href{https://arxiv.org/abs/1104.1564}{{arXiv:1104.1564}}}
{[hep-ph]}
\end{barticle}
\endbibitem

\bibitem[\protect\citeauthoryear{Alkofer and von Smekal}{2001}]{Alkofer:2000wg}
\begin{barticle}
\bauthor{\bsnm{Alkofer}, \binits{R.}},
\bauthor{\bsnm{Smekal}, \binits{L.}}:
\batitle{{The Infrared behavior of QCD Green's functions: Confinement dynamical
  symmetry breaking, and hadrons as relativistic bound states}}.
\bjtitle{Phys. Rept.}
\bvolume{353},
\bfpage{281}
(\byear{2001})
\doiurl{10.1016/S0370-1573(01)00010-2}
{\href{https://arxiv.org/abs/hep-ph/0007355}{{arXiv:hep-ph/0007355}}}
\end{barticle}
\endbibitem

\bibitem[\protect\citeauthoryear{Roberts}{2008}]{Roberts:2007ji}
\begin{barticle}
\bauthor{\bsnm{Roberts}, \binits{C.D.}}:
\batitle{{Hadron Properties and Dyson-Schwinger Equations}}.
\bjtitle{Prog. Part. Nucl. Phys.}
\bvolume{61},
\bfpage{50}--\blpage{65}
(\byear{2008})
\doiurl{10.1016/j.ppnp.2007.12.034}
{\href{https://arxiv.org/abs/0712.0633}{{arXiv:0712.0633}}}
{[nucl-th]}
\end{barticle}
\endbibitem

\bibitem[\protect\citeauthoryear{Fischer et~al.}{2014}]{Fischer:2014ata}
\begin{barticle}
\bauthor{\bsnm{Fischer}, \binits{C.S.}},
\bauthor{\bsnm{Luecker}, \binits{J.}},
\bauthor{\bsnm{Welzbacher}, \binits{C.A.}}:
\batitle{{Phase structure of three and four flavor QCD}}.
\bjtitle{Phys. Rev. D}
\bvolume{90}(\bissue{3}),
\bfpage{034022}
(\byear{2014})
\doiurl{10.1103/PhysRevD.90.034022}
{\href{https://arxiv.org/abs/1405.4762}{{arXiv:1405.4762}}}
{[hep-ph]}
\end{barticle}
\endbibitem

\bibitem[\protect\citeauthoryear{Gao et~al.}{2021}]{Gao:2021wun}
\begin{barticle}
\bauthor{\bsnm{Gao}, \binits{F.}},
\bauthor{\bsnm{Papavassiliou}, \binits{J.}},
\bauthor{\bsnm{Pawlowski}, \binits{J.M.}}:
\batitle{{Fully coupled functional equations for the quark sector of QCD}}.
\bjtitle{Phys. Rev. D}
\bvolume{103}(\bissue{9}),
\bfpage{094013}
(\byear{2021})
\doiurl{10.1103/PhysRevD.103.094013}
{\href{https://arxiv.org/abs/2102.13053}{{arXiv:2102.13053}}}
{[hep-ph]}
\end{barticle}
\endbibitem

\bibitem[\protect\citeauthoryear{Vija and Thoma}{1995}]{Vija:1994is}
\begin{barticle}
\bauthor{\bsnm{Vija}, \binits{H.}},
\bauthor{\bsnm{Thoma}, \binits{M.H.}}:
\batitle{{Braaten-Pisarski method at finite chemical potential}}.
\bjtitle{Phys. Lett. B}
\bvolume{342},
\bfpage{212}--\blpage{218}
(\byear{1995})
\doiurl{10.1016/0370-2693(94)01378-P}
{\href{https://arxiv.org/abs/hep-ph/9409246}{{arXiv:hep-ph/9409246}}}
\end{barticle}
\endbibitem

\bibitem[\protect\citeauthoryear{Haque et~al.}{2013}]{Haque:2012my}
\begin{barticle}
\bauthor{\bsnm{Haque}, \binits{N.}},
\bauthor{\bsnm{Mustafa}, \binits{M.G.}},
\bauthor{\bsnm{Strickland}, \binits{M.}}:
\batitle{{Two-loop hard thermal loop pressure at finite temperature and
  chemical potential}}.
\bjtitle{Phys. Rev. D}
\bvolume{87}(\bissue{10}),
\bfpage{105007}
(\byear{2013})
\doiurl{10.1103/PhysRevD.87.105007}
{\href{https://arxiv.org/abs/1212.1797}{{arXiv:1212.1797}}}
{[hep-ph]}
\end{barticle}
\endbibitem

\bibitem[\protect\citeauthoryear{Bowman et~al.}{2005}]{Bowman:2005vx}
\begin{barticle}
\bauthor{\bsnm{Bowman}, \binits{P.O.}},
\bauthor{\bsnm{Heller}, \binits{U.M.}},
\bauthor{\bsnm{Leinweber}, \binits{D.B.}},
\bauthor{\bsnm{Parappilly}, \binits{M.B.}},
\bauthor{\bsnm{Williams}, \binits{A.G.}},
\bauthor{\bsnm{Zhang}, \binits{J.-b.}}:
\batitle{{Unquenched quark propagator in Landau gauge}}.
\bjtitle{Phys. Rev. D}
\bvolume{71},
\bfpage{054507}
(\byear{2005})
\doiurl{10.1103/PhysRevD.71.054507}
{\href{https://arxiv.org/abs/hep-lat/0501019}{{arXiv:hep-lat/0501019}}}
\end{barticle}
\endbibitem

\bibitem[\protect\citeauthoryear{Gao and Liu}{2016}]{Gao:2016qkh}
\begin{barticle}
\bauthor{\bsnm{Gao}, \binits{F.}},
\bauthor{\bsnm{Liu}, \binits{Y.-x.}}:
\batitle{{QCD phase transitions via a refined truncation of Dyson-Schwinger
  equations}}.
\bjtitle{Phys. Rev. D}
\bvolume{94}(\bissue{7}),
\bfpage{076009}
(\byear{2016})
\doiurl{10.1103/PhysRevD.94.076009}
{\href{https://arxiv.org/abs/1607.01675}{{arXiv:1607.01675}}}
{[hep-ph]}
\end{barticle}
\endbibitem

\bibitem[\protect\citeauthoryear{Lu et~al.}{2023}]{Lu:2023mkn}
\begin{botherref}
\oauthor{\bsnm{Lu}, \binits{Y.}},
\oauthor{\bsnm{Gao}, \binits{F.}},
\oauthor{\bsnm{Liu}, \binits{Y.-X.}},
\oauthor{\bsnm{Pawlowski}, \binits{J.M.}}:
{QCD equation of state and thermodynamic observables from computationally
  minimal Dyson-Schwinger Equations}
(2023)
{\href{https://arxiv.org/abs/2310.18383}{{arXiv:2310.18383}}}
{[hep-ph]}
\end{botherref}
\endbibitem

\bibitem[\protect\citeauthoryear{Bernhardt and
  Fischer}{2023}]{Bernhardt:2023ezo}
\begin{barticle}
\bauthor{\bsnm{Bernhardt}, \binits{J.}},
\bauthor{\bsnm{Fischer}, \binits{C.S.}}:
\batitle{{From imaginary to real chemical potential QCD with functional
  methods}}.
\bjtitle{Eur. Phys. J. A}
\bvolume{59}(\bissue{8}),
\bfpage{181}
(\byear{2023})
\doiurl{10.1140/epja/s10050-023-01098-1}
{\href{https://arxiv.org/abs/2305.01434}{{arXiv:2305.01434}}}
{[hep-ph]}
\end{barticle}
\endbibitem

\bibitem[\protect\citeauthoryear{Basar}{2021}]{Basar:2021hdf}
\begin{barticle}
\bauthor{\bsnm{Basar}, \binits{G.}}:
\batitle{{Universality, Lee-Yang Singularities, and Series Expansions}}.
\bjtitle{Phys. Rev. Lett.}
\bvolume{127}(\bissue{17}),
\bfpage{171603}
(\byear{2021})
\doiurl{10.1103/PhysRevLett.127.171603}
{\href{https://arxiv.org/abs/2105.08080}{{arXiv:2105.08080}}}
{[hep-th]}
\end{barticle}
\endbibitem

\bibitem[\protect\citeauthoryear{Zinn-Justin}{2021}]{Zinn-Justin:1989rgp}
\begin{bbook}
\bauthor{\bsnm{Zinn-Justin}, \binits{J.}}:
\bbtitle{{Quantum Field Theory and Critical Phenomena}}.
\bsertitle{International Series of Monographs on Physics},
vol. \bseriesno{77}.
\bpublisher{Oxford University Press}, \blocation{???}
(\byear{2021})
\end{bbook}
\endbibitem

\bibitem[\protect\citeauthoryear{Connelly et~al.}{2020}]{Connelly:2020gwa}
\begin{barticle}
\bauthor{\bsnm{Connelly}, \binits{A.}},
\bauthor{\bsnm{Johnson}, \binits{G.}},
\bauthor{\bsnm{Rennecke}, \binits{F.}},
\bauthor{\bsnm{Skokov}, \binits{V.}}:
\batitle{{Universal Location of the Yang-Lee Edge Singularity in $O(N)$
  Theories}}.
\bjtitle{Phys. Rev. Lett.}
\bvolume{125}(\bissue{19}),
\bfpage{191602}
(\byear{2020})
\doiurl{10.1103/PhysRevLett.125.191602}
{\href{https://arxiv.org/abs/2006.12541}{{arXiv:2006.12541}}}
{[cond-mat.stat-mech]}
\end{barticle}
\endbibitem

\bibitem[\protect\citeauthoryear{Johnson et~al.}{2023}]{Johnson:2022cqv}
\begin{barticle}
\bauthor{\bsnm{Johnson}, \binits{G.}},
\bauthor{\bsnm{Rennecke}, \binits{F.}},
\bauthor{\bsnm{Skokov}, \binits{V.V.}}:
\batitle{{Universal location of Yang-Lee edge singularity in classic O(N)
  universality classes}}.
\bjtitle{Phys. Rev. D}
\bvolume{107}(\bissue{11}),
\bfpage{116013}
(\byear{2023})
\doiurl{10.1103/PhysRevD.107.116013}
{\href{https://arxiv.org/abs/2211.00710}{{arXiv:2211.00710}}}
{[hep-ph]}
\end{barticle}
\endbibitem

\bibitem[\protect\citeauthoryear{Rennecke and Skokov}{2022}]{Rennecke:2022ohx}
\begin{barticle}
\bauthor{\bsnm{Rennecke}, \binits{F.}},
\bauthor{\bsnm{Skokov}, \binits{V.V.}}:
\batitle{{Universal location of Yang\textendash{}Lee edge singularity for a
  one-component field theory in 1\ensuremath{\leq}d\ensuremath{\leq}4}}.
\bjtitle{Annals Phys.}
\bvolume{444},
\bfpage{169010}
(\byear{2022})
\doiurl{10.1016/j.aop.2022.169010}
{\href{https://arxiv.org/abs/2203.16651}{{arXiv:2203.16651}}}
{[hep-ph]}
\end{barticle}
\endbibitem

\bibitem[\protect\citeauthoryear{Kos et~al.}{2016}]{Kos:2016ysd}
\begin{barticle}
\bauthor{\bsnm{Kos}, \binits{F.}},
\bauthor{\bsnm{Poland}, \binits{D.}},
\bauthor{\bsnm{Simmons-Duffin}, \binits{D.}},
\bauthor{\bsnm{Vichi}, \binits{A.}}:
\batitle{{Precision Islands in the Ising and $O(N)$ Models}}.
\bjtitle{JHEP}
\bvolume{08},
\bfpage{036}
(\byear{2016})
\doiurl{10.1007/JHEP08(2016)036}
{\href{https://arxiv.org/abs/1603.04436}{{arXiv:1603.04436}}}
{[hep-th]}
\end{barticle}
\endbibitem

\bibitem[\protect\citeauthoryear{Kaczmarek et~al.}{2011}]{Kaczmarek:2011zz}
\begin{barticle}
\bauthor{\bsnm{Kaczmarek}, \binits{O.}},
\bauthor{\bsnm{Karsch}, \binits{F.}},
\bauthor{\bsnm{Laermann}, \binits{E.}},
\bauthor{\bsnm{Miao}, \binits{C.}},
\bauthor{\bsnm{Mukherjee}, \binits{S.}},
\bauthor{\bsnm{Petreczky}, \binits{P.}},
\bauthor{\bsnm{Schmidt}, \binits{C.}},
\bauthor{\bsnm{Soeldner}, \binits{W.}},
\bauthor{\bsnm{Unger}, \binits{W.}}:
\batitle{{Phase boundary for the chiral transition in (2+1) -flavor QCD at
  small values of the chemical potential}}.
\bjtitle{Phys. Rev. D}
\bvolume{83},
\bfpage{014504}
(\byear{2011})
\doiurl{10.1103/PhysRevD.83.014504}
{\href{https://arxiv.org/abs/1011.3130}{{arXiv:1011.3130}}}
{[hep-lat]}
\end{barticle}
\endbibitem

\bibitem[\protect\citeauthoryear{Mukherjee and
  Skokov}{2021}]{Mukherjee:2019eou}
\begin{barticle}
\bauthor{\bsnm{Mukherjee}, \binits{S.}},
\bauthor{\bsnm{Skokov}, \binits{V.}}:
\batitle{{Universality driven analytic structure of the QCD crossover: radius
  of convergence in the baryon chemical potential}}.
\bjtitle{Phys. Rev. D}
\bvolume{103}(\bissue{7}),
\bfpage{071501}
(\byear{2021})
\doiurl{10.1103/PhysRevD.103.L071501}
{\href{https://arxiv.org/abs/1909.04639}{{arXiv:1909.04639}}}
{[hep-ph]}
\end{barticle}
\endbibitem

\bibitem[\protect\citeauthoryear{Goswami et~al.}{2024}]{Goswami:2024jlc}
\begin{botherref}
\oauthor{\bsnm{Goswami}, \binits{J.}},
\oauthor{\bsnm{Clarke}, \binits{D.A.}},
\oauthor{\bsnm{Dimopoulos}, \binits{P.}},
\oauthor{\bsnm{Di~Renzo}, \binits{F.}},
\oauthor{\bsnm{Schmidt}, \binits{C.}},
\oauthor{\bsnm{Singh}, \binits{S.}},
\oauthor{\bsnm{Zambello}, \binits{K.}}:
{Exploring the Critical Points in QCD with Multi-Point Pad\'e and Machine
  Learning Techniques in (2+1)-flavor QCD}
(2024)
{\href{https://arxiv.org/abs/2401.05651}{{arXiv:2401.05651}}}
{[hep-lat]}
\end{botherref}
\endbibitem

\bibitem[\protect\citeauthoryear{Dimopoulos et~al.}{2022}]{Dimopoulos:2022}
\begin{barticle}
\bauthor{\bsnm{Dimopoulos}, \binits{P.}},
\bauthor{\bsnm{Dini}, \binits{L.}},
\bauthor{\bsnm{Di~Renzo}, \binits{F.}},
\bauthor{\bsnm{Goswami}, \binits{J.}},
\bauthor{\bsnm{Nicotra}, \binits{G.}},
\bauthor{\bsnm{Schmidt}, \binits{C.}},
\bauthor{\bsnm{Singh}, \binits{S.}},
\bauthor{\bsnm{Zambello}, \binits{K.}},
\bauthor{\bsnm{Ziesch\'e}, \binits{F.}}:
\batitle{Contribution to understanding the phase structure of strong
  interaction matter: Lee-yang edge singularities from lattice qcd}.
\bjtitle{Phys. Rev. D}
\bvolume{105},
\bfpage{034513}
(\byear{2022})
\doiurl{10.1103/PhysRevD.105.034513}
\end{barticle}
\endbibitem

\bibitem[\protect\citeauthoryear{Bollweg et~al.}{2022}]{Bollweg:2022rps}
\begin{barticle}
\bauthor{\bsnm{Bollweg}, \binits{D.}},
\bauthor{\bsnm{Goswami}, \binits{J.}},
\bauthor{\bsnm{Kaczmarek}, \binits{O.}},
\bauthor{\bsnm{Karsch}, \binits{F.}},
\bauthor{\bsnm{Mukherjee}, \binits{S.}},
\bauthor{\bsnm{Petreczky}, \binits{P.}},
\bauthor{\bsnm{Schmidt}, \binits{C.}},
\bauthor{\bsnm{Scior}, \binits{P.}}:
\batitle{{Taylor expansions and Pad\'e approximants for cumulants of conserved
  charge fluctuations at nonvanishing chemical potentials}}.
\bjtitle{Phys. Rev. D}
\bvolume{105}(\bissue{7}),
\bfpage{074511}
(\byear{2022})
\doiurl{10.1103/PhysRevD.105.074511}
{\href{https://arxiv.org/abs/2202.09184}{{arXiv:2202.09184}}}
{[hep-lat]}
\end{barticle}
\endbibitem

\bibitem[\protect\citeauthoryear{Stephanov}{2006}]{Stephanov:2006}
\begin{barticle}
\bauthor{\bsnm{Stephanov}, \binits{M.A.}}:
\batitle{Qcd critical point and complex chemical potential singularities}.
\bjtitle{Phys. Rev. D}
\bvolume{73},
\bfpage{094508}
(\byear{2006})
\doiurl{10.1103/PhysRevD.73.094508}
\end{barticle}
\endbibitem

\bibitem[\protect\citeauthoryear{Lu et~al.}{2023}]{Lu:2023msn}
\begin{botherref}
\oauthor{\bsnm{Lu}, \binits{Y.}},
\oauthor{\bsnm{Gao}, \binits{F.}},
\oauthor{\bsnm{Fu}, \binits{B.-C.}},
\oauthor{\bsnm{Song}, \binits{H.-C.}},
\oauthor{\bsnm{Liu}, \binits{Y.-X.}}:
{Constructing the Equation of State of QCD in a functional QCD based scheme}
(2023)
{\href{https://arxiv.org/abs/2310.16345}{{arXiv:2310.16345}}}
{[hep-ph]}
\end{botherref}
\endbibitem

\bibitem[\protect\citeauthoryear{Parotto et~al.}{2020}]{Parotto:2020}
\begin{barticle}
\bauthor{\bsnm{Parotto}, \binits{P.}},
\bauthor{\bsnm{Bluhm}, \binits{M.}},
\bauthor{\bsnm{Mroczek}, \binits{D.}},
\bauthor{\bsnm{Nahrgang}, \binits{M.}},
\bauthor{\bsnm{Noronha-Hostler}, \binits{J.}},
\bauthor{\bsnm{Rajagopal}, \binits{K.}},
\bauthor{\bsnm{Ratti}, \binits{C.}},
\bauthor{\bsnm{Sch\"afer}, \binits{T.}},
\bauthor{\bsnm{Stephanov}, \binits{M.}}:
\batitle{Qcd equation of state matched to lattice data and exhibiting a
  critical point singularity}.
\bjtitle{Phys. Rev. C}
\bvolume{101},
\bfpage{034901}
(\byear{2020})
\doiurl{10.1103/PhysRevC.101.034901}
\end{barticle}
\endbibitem

\bibitem[\protect\citeauthoryear{Rehr and Mermin}{1973}]{Rehr:1973zz}
\begin{barticle}
\bauthor{\bsnm{Rehr}, \binits{J.J.}},
\bauthor{\bsnm{Mermin}, \binits{N.D.}}:
\batitle{Revised scaling equation of state at the liquid-vapor critical point}.
\bjtitle{Phys. Rev. A}
\bvolume{8},
\bfpage{472}--\blpage{480}
(\byear{1973})
\doiurl{10.1103/PhysRevA.8.472}
\end{barticle}
\endbibitem

\bibitem[\protect\citeauthoryear{Braun et~al.}{2023}]{fQCD}
\begin{botherref}
\oauthor{\bsnm{Braun}, \binits{J.}},
\oauthor{\bsnm{Chen}, \binits{Y.-r.}},
\oauthor{\bsnm{Fu}, \binits{W.-j.}},
\oauthor{\bsnm{Gao}, \binits{F.}},
\oauthor{\bsnm{Geissel}, \binits{A.}},
\oauthor{\bsnm{Horak}, \binits{J.}},
\oauthor{\bsnm{Huang}, \binits{C.}},
\oauthor{\bsnm{Ihssen}, \binits{F.}},
\oauthor{\bsnm{Lu}, \binits{Y.}},
\oauthor{\bsnm{Pawlowski}, \binits{J.M.}},
\oauthor{\bsnm{Rennecke}, \binits{F.}},
\oauthor{\bsnm{Sattler}, \binits{F.}},
\oauthor{\bsnm{Schallmo}, \binits{B.}},
\oauthor{\bsnm{Stoll}, \binits{J.}},
\oauthor{\bsnm{Tan}, \binits{Y.-y.}},
\oauthor{\bsnm{T{\"o}pfel}, \binits{S.}},
\oauthor{\bsnm{Turnwald}, \binits{J.}},
\oauthor{\bsnm{Wen}, \binits{R.}},
\oauthor{\bsnm{Wessely}, \binits{J.}},
\oauthor{\bsnm{Wink}, \binits{N.}},
\oauthor{\bsnm{Yin}, \binits{S.}},
\oauthor{\bsnm{Zorbach}, \binits{N.}}:
{fQCD collaboration}
(2023)
\end{botherref}
\endbibitem

\end{thebibliography}

\end{sloppypar}

\end{document}